\documentclass[aps,prd,superscriptaddress,showpacs,preprintnumbers]{revtex4}

\usepackage{color}
\usepackage{epsfig}
\usepackage{graphicx,graphics}
\usepackage{amsmath}
\usepackage{amsfonts}
\usepackage{amssymb}
\usepackage{multirow}

\newcommand{\mc}{\mathcal}

\newcommand{\bi}{\begin{itemize}}
\newcommand{\ei}{\end{itemize}}

\usepackage{bm}

\newcommand{\nl}{\newline}
\newcommand{\beq}{\begin{equation}}
 \newcommand{\eeq}{\end{equation}}

 \newcommand{\be}{\begin{eqnarray}}
 \newcommand{\ee}{\end{eqnarray}}
 \newcommand{\nee}{\nonumber\end{eqnarray}}
 \newcommand{\nn}{\nonumber\\}

  \newcommand{\bc}{\begin{center}}
 \newcommand{\ec}{\end{center}}

\def\kt{k_\perp}

\def\pp{p_\perp}

\def\avk{\langle k_\perp ^2\rangle}
\def\avp{\langle p_\perp ^2\rangle}
\def\avPT{\langle P_T^2\rangle}

\def\J{_{_J}}
\def\S{_{_S}}

\def\C{_{_C}}

\def\BM{_{_{B\!M}}}

\def\xb{x_{_{\!B}}}
\def\1{_{_{C\!1}}}
\def\2{_{_{S1}}}

  \def\a               {\alpha}
\def\b               {\beta}

\def\s              {\sigma}
\def\g              {\gamma}

\begin{document}

\author{E. Christova}
\email {echristo@inrne.bas.bg}
\affiliation{Institute for Nuclear Research and Nuclear Energy, Bulgarian Academy of Sciences, Tzarigradsko chauss\'{e}e 72, 1784 Sofia, Bulgaria}
\author{E. Leader}
\email {elliot.leader@cern.ch}
\affiliation{ Imperial College London, London SW7 2AZ, United Kingdom}
\author{M. Stoilov}
\email {mstoilov@inrne.bas.bg}
\affiliation{Institute for Nuclear Research and Nuclear Energy, Bulgarian Academy of Sciences, Tzarigradsko chauss\'{e}e 72, 1784 Sofia, Bulgaria}

\title{Consistency tests  for the extraction of the Boer-Mulders  and Sivers functions}

\begin{abstract}
At present, the Boer-Mulders (BM) function for a given quark flavour is extracted from data on semi-inclusive deep inelastic scattering (SIDIS)
using  the simplifying assumption that it is proportional to the Sivers function for that flavour. In a recent paper we suggested that
 the consistency of this assumption could be tested using information on so-called difference asymmetries
  i.e. the difference
  between the asymmetries  in the production of particles and their anti-particles.

    In this paper, using the SIDIS COMPASS deuteron data on
the $\langle\cos\phi_h\rangle$,  $\langle\cos 2 \phi_h\rangle$ and
Sivers difference asymmetries, we carry out  two  independent consistency
tests of the assumption of proportionality, but here applied to the sum of the valence-quark
contributions.
We find that such an assumption is compatible with  the data. We also show that the proportionality assumptions made in the existing parametrizations  of the BM functions are not compatible with our analysis, which suggests that the published results for the Boer-Mulders functions for individual flavours are unreliable.

    %This
  % suggests that the  published information on the BM functions
   %   for individual quark flavours may be unreliable.

The $\langle\cos\phi_h\rangle$ and   $\langle\cos 2 \phi_h\rangle$ asymmetries
receive contributions also from the, in principle, calculable Cahn
effect. We succeed in extracting the Cahn contributions  from experiment (we believe for the first time) and compare with their calculated
values, with interesting implications.

\end{abstract}

\pacs{...}

\date{\today}

\maketitle
\bc

\ec

\section{Introduction}

There is a major  effort at present to progress beyond a knowledge of collinear  parton distribution functions (PDFs) and fragmentation functions (FFs) and to obtain information about  the transverse momentum dependent (TMD)
versions of these functions. In extracting these distributions from data  a standard
parametrization  is usually adopted (see for example \cite{general}), which involves various simplifying assumptions.
In addition, because of lack of sufficient data,  additional relations between different TMD-functions are sometimes assumed.
We focus on, and  examine, the particular  assumption that the BM functions for a particular flavour
are proportional to the Sivers functions of the same flavour.\newline

In our recent paper \cite{we} we showed that the difference asymmetries in SIDIS allow the determination of the valence quark TMDs in a model independent way, without
any assumptions about the sea quark  or gluon densities.  Also, that using the difference asymmetries, one can test many of
 the basic assumptions in the standard parametrization, such as factorization of the $\xb$- and $z_h$-dependencies,
 the Gaussian flavour- and hadron-independent
$k_\perp$-behaviour etc.

In ~\cite{we} we derived two types of relations -- between the $\langle\cos\phi_h\rangle$, $\langle\cos 2 \phi_h\rangle$ and Sivers asymmetries, that allow  tests of the simplifying assumption used in extracting the Boer-Mulders (BM) function i.e. its proportionality to the Sivers function~\cite{BM_1,BM_2}, an assumption  motivated by  model calculations~\cite{model}. In addition, present analyses make a further assumption concerning the $Q^2$ evolution of these functions for  a given quark flavor, which, as explained in the next Section, is theoretically inconsistent.

 Our previously published tests~\cite{we} were formulated without taking into account the Cahn effect, which inevitably contributes to these asymmetries.
In this paper we show how these tests are modified when the Cahn effect  is included.

 We then use  COMPASS SIDIS measurements of
the $\langle\cos\phi_h\rangle$,  $\langle\cos 2 \phi_h\rangle$ and Sivers asymmetries on a deuteron target
to  test for the consistency of the assumed relation between BM and Sivers functions.

  We work with the so called difference asymmetries  of  the following general structure. If the asymmetries for $h^+$ and $h^-$ have the form
  \be
  A^{h^+}= \frac{\Delta \sigma^{h^+}}{\sigma^{h^+}} \qquad A^{h^-}= \frac{\Delta \sigma^{h^-}}{\sigma^{h^-}},
   \ee
  where $\s^{h^+, \,h^-}$   and $\Delta \s^{h^+, \,h^-}$ are  the  unpolarized and polarized  cross sections respectively, then
 \be
  A^{h^+ - h^-}\equiv \frac{\Delta \s^{h^+}-\Delta \s^{ h^-}}{\s^{h^+} -\s^{ h^-}}.\label{A}
\ee

 The difference asymmetries are expressed in terms of the usual  asymmetries $ A^{h^+,\, h^-}$
 and the ratio of the corresponding multiplicities \cite{COMPASS-diff}:

 \beq
    A^{h^+ -h^-} = \frac{1}{1-r} \left( A^{h^+} - r A^{h^-}\right) .
    \label{eq.Diff1}
     \eeq
where $r$ is the ratio of unpolarized  SIDIS
 cross sections for production of  $h^-$ and $h^+$:
 $ r = \sigma^{h^-} / \sigma^{h^+}$.

  As shown in ref.\cite{we},  the advantage of using the difference asymmetries is that,
   based only on charge conjugation
(C) and isospin (SU(2)) invariance of the strong interactions,
   they are expressed  purely in terms of the best known valence-quark
    distributions and  fragmentation functions;
 sea-quark and gluon distributions do not enter. For a deuteron target there is the additional simplification that,
 independently of the final hadron,  only the sum of the valence-quark distributions   enters.\\

 The paper is organized as follows: the  notation and conventions for the various TMD
 functions and the used experimental asymmetries are explained in Sections II and III; in Section IV
 we formulate the two tests for
 the assumed relation between the BM and Sivers functions.
  They are based  on the $\langle\cos\phi_h\rangle$ and  $\langle\cos 2\phi_h\rangle$
   azimuthal asymmetries  of the final hadrons in unpolarized SIDIS,
 and the Sivers asymmetry for unpolarized leptons on transversely polarized nucleons.
 Because the above two unpolarized asymmetries receive contributions from both the BM and Cahn effects,
 we are able also to extract information about the Cahn effect; in Section V
   we apply these tests using the COMPASS SIDIS data on deuterons.\\

%%%%%%%%%%%%%%%%%%%%%%%%%%%%%%%%%%%%%%%%%%%%%%%%%%%%%%%%%%%%%%%%%%%%%%%%%%%%%%%%%%%%%%%%%%%%%%%%%%%%%%
\section{Parametrization of the TMD distributions}

%%%%%%%%%%%%%%%%%%%%%%%%%%%%%%%%%%%%%%%%%%%%%%%%%%%%%%%%%%%%%%%%%%%%%%%%%%%%%%%%%%%%%%%%%
\subsection{The polarized parton distribution functions}

Conventionally,  a typical spin-dependent  TMD  density $\Delta f(k_\perp ,\xb , Q^2)$  has been  parametrized following
several simplifying assumptions:

1) The transverse-momentum dependence  on $k_\perp $ is
factorized from the $\xb $-dependence.

2) The $k_\perp $-dependence  is flavour and hadron independent, and usually assumed to be a Gaussian.

We adopt these two simplifications.

 3)   An additional simplifying
   assumption  is that TMD functions are proportional to
   the related
    collinear parton distribution functions (PDFs) and   fragmentation functions
    (FFs). The $Q^2$-evolution is usually  assumed  to be given via the collinear PDFs and FFs, i.e.  making the  ansatz:
 \be
\Delta f_q(\xb ,Q^2)&=&2\,{\cal N}_q(\xb ) q(\xb ,Q^2)\nn \Delta
f_{\bar q}(\xb ,Q^2)&=&2\,{\cal N}_{\bar q}(\xb ) \bar q(\xb
,Q^2)\label{TMD-q}
\ee
This
is, however,  physically unacceptable because it leads to gluons contributing to the evolution of non-singlet combinations
of quark densities.

Since we deal here only with valence quark densities  we replace this simplification by  an ansatz for the valence-quark densities.
 Hence we take the $Q^2$ evolution to be controlled via:
\be
\Delta f_{q_V}(\xb ,Q^2)=2\,{\cal N}_{q_V}(\xb ) q_V(\xb ,Q^2),\qquad q_V=u_V,d_V\label{TMD-qV}
\ee
Note, however, that we do not think this difference in approximating the evolution is important
 when assessing the impact   of our tests  on the published BM data.

In this paper we consider only the difference asymmetries on a deuteron target. As mentioned earlier, in these asymmetries only
one combination of parton density enters -- the sum of the valence-quark TMD functions:
\be
\Delta f_{Q_V} (\xb,\kt,Q^2)\equiv \Delta f_{u_V} (\xb,\kt,Q^2)+\Delta f_{d_V} (\xb,\kt,Q^2)
\ee

  Below we present the parametrizations of the valence-quark $Q_V$ unpolarized, BM and Sivers distributions and
 the Collins FFs following the above simplifying anzatz.
 We work in the approximation ${\cal O}(k_\perp/Q)$, neglecting terms of the order ${\cal O}(k_\perp^2/Q^2)$.

 %%%%%%%%%%%%%%%%%%%%%%%%%%%%%%%%%%%%%%%%%%%%%%%%%%%%%%%%%%%%%%%%%%%%%%%%%%%%%%%%%%%%%%%%%%%%%%%%%%%%%%%%%%%%%%%%%%%%%%%%%%%%%%

\subsection{The unpolarized TMD parton distributions and fragmentation functions.}\label{TMD-unpol}

 The unpolarized TMD PDFs and FFs  are
parametrized proportional to the corresponding collinear functions times a Gaussian-type, flavour and hadron independent
$k_\perp^2/p_\perp^2$ dependence~\cite{TMD}. In accordance with this for
the valence-quark unpolarized TMD PDFs $f_{Q_V/p}(\xb, k_\perp^2, Q^2)$ and TMD FFs $D_{h/q_V}(z_h, p_\perp^2, Q^2)$
 we adopt the parametrizations~\cite{we0}:
\be
f_{Q_V/p}(\xb,k_\perp^2 ,
Q^2)&=&Q_V(\xb,Q^2)\,\frac{e^{-k_\perp^2/\avk}}{\pi\avk}\label{fq}
\ee
and
 \be
D_{h/q_V}(z_h,p_\perp^2 ,Q^2) =D_{q_V}^h(z_h,Q^2)
\,\frac{e^{-p_\perp^2/\avp }}{\pi\avp },\label{Dq}
\ee
where $Q_V(\xb,Q^2)$ is the sum of the collinear valence-quark  PDFs:
 \be
 Q_V(\xb,Q^2)= u_V(\xb,Q^2) +d_V(\xb,Q^2)
  \ee
  and $D_{q_V}^h(z_h,Q^2)$ are the valence-quark collinear  FFs:
  \be
  D_{q_V}^h(z_h,Q^2)=D_{q}^h(z_h,Q^2)-D_{\bar q}^h(z_h,Q^2),
  \ee
   $\avk $ and $\avp$ are
parameters extracted from study of the multiplicities in unpolarized SIDIS.

The parameters  $\avk$ (and $\avp$  are basic  as they enter in the  normalization functions in all TMD asymmetries.
At present the experimentally obtained values are controversial:

 1)  $\avk \approx 0.25\,GeV^2$ and $\avp \approx 0.20\,GeV^2$~\cite{Anselmino_2005}, extracted from the
  old EMC~\cite{EMC}  and FNAL~\cite{FNAL} SIDIS data

  2) $\avk = 0.18\,GeV^2$ and $\avp = 0.20\,GeV^2$ \cite{MonteCarlo}, derived from the $P_T$-spectrum of
  HERMES data and confirmed by Monte Carlo calculations.
  The extraction of the BM functions in \cite{BM_2} utilized these values.

An analysis \cite{TMD} of the more recent available data on multiplicities from HERMES~\cite{HERMES-m} and
COMPASS~\cite{COMPASS-m} separately,  gives quite different values:

3) $\avk = 0.57 \pm 0.08\,GeV^2$ and $\avp = 0.12\pm 0.01\,GeV^2$, extracted from  HERMES data

4) $\avk = 0.61 \pm 0.20\,GeV^2$ and $\avp = 0.19\pm 0.02\,GeV^2$, extracted from  COMPASS data.

These values are obtained using a kinematical cut on $z_h<0.6$ and
 they  change slightly on placing the  cut at $z_h<0.7$.

  Further we shall be able to comment on this controversial situation,
 since the Cahn effect, which contributes to the asymmetries which we study and  extract from data,
  is  calculable, and depends sensitively on $\avk$ and $\avp$.

%%%%%%%%%%%%%%%%%%%%%%%%%%%%%%%%%%%%%%%%%%%%%%%%%%%%%%%%%%%%%%%%%%%%%%%%%%%%%%%%%%%%%%%%%%%%%%%%%%%%%%%%%%%%%%%%%%%%%%%%%%%%%%

\subsection{The BM and Sivers distributions}

The Sivers function describes the correlation between the spin of the nucleon $\bf S$,  its
momentum $\bf{P}$,  and the momentum of the quark ${\bf k}_\perp$, via a term proportional to  $\bf { S \cdot (k_\perp \times  P) }$ ~\cite{Sivers},
while  the BM function describes the correlation between the spin of the quark ${\bf s}_q$
and the momentum of the quark ${\bf k}_\perp$, via a term  proportional to
  $ \bf { s_q \cdot (k_\perp \times  P) }$ ~\cite{BM}.

The $k_\bot, x_B$ dependence of the valence-quark BM and Sivers distribution functions $\Delta f_J^{Q_V}(\xb,\kt ,Q^2)$, ($J=$ BM, Sivers),
is assumed to factorize ~\cite{general,BM_2} in the form
\beq
\hspace*{-.5cm}  \Delta  f^{Q_V}_J(\xb,\kt ,Q^2) \!=\! \Delta  f^{Q_V}\J(\xb,Q^2)\; \sqrt{2e}\,\frac{\kt}{M\J} \;
\frac{e^{-\kt^2/\avk\J }}{\pi\avk\J },\qquad J=BM, Sivers\label{BM-Siv_dist1} \eeq
with
\beq\Delta
f^{Q_V}\J(\xb,Q^2)\!=\! 2\,{\cal N}\J^{Q_V}(\xb)\,Q_V(\xb,Q^2) \label{BM-Siv_dist2}
 \eeq
Here  the ${\cal N}^{Q_V}\J(\xb)$ are unknown functions, and $M\J$,  or equivalently $\avk \J$, where
\be
 \avk \J= \frac{\avk  \, M^2\J}{\avk  + M^2 \J},
 \ee
are unknown  parameters. As mentioned earlier,  $\avk$ is supposed to be known from multiplicities in unpolarized SIDIS.

%%%%%%%%%%%%%%%%%%%%%%%%%%%%%%%%%%%%%%%%%%%%%%%%%%%%%%%%%%%%%%%%%%%%%%%%%%%%%%%%%%%%%%%%%%%%%%%%%%%%%%%
\subsection{The Collins fragmentation functions}

The Collins fragmentation functions (FFs) $\Delta^N
D_{h/q\uparrow}(z,\pp)$ describe phenomenologically the spin-dependent part of the fragmentation functions
of transversely polarized quarks, with transverse spin  ${\bf  s}_q$ and
3-momentum  ${\bf  p}_q$, into hadrons $h$ with momentum ${\bf p}_\perp$, transverse to the direction of the initial quark~\cite{Collins}:
 \be
D_{h/q,s}(z_h,\pp)=D_{h/q}(z_h,\pp)+ \frac{1}{2}\, \Delta^N
D_{h/q\uparrow}(z_h,\pp)\,{\bf \hat s}_q\cdot({\bf \hat p}_q\times {\bf
\hat p}_\perp).
 \ee
 It relates the transverse momentum of the produced hadron to the transverse spin of the quark and  and leads to nonuniform azimuthal distribution of
final hadrons around the initial quark direction.

  The valence-quark Collins functions $ \Delta^N  D_{h/u_V\uparrow}(z_h,\pp ,Q^2) $ are parametrized \cite{we} proportional to the
corresponding unpolarized valence-quark collinear fragmentation
functions $D_{u_V}^h(z_h,Q^2) $:
\beq
 \Delta^N  D_{h/u_V\uparrow}(z_h,\pp ,Q^2) \!=\!\Delta^N  D_{h/u_V\uparrow}(z_h,Q^2)\,
\sqrt{2e}\,\frac{\pp}{M\C} \; \frac{e^{-\pp^2/\avp\C }}{\pi\avp\C}\,,\quad h=\pi^+,K^+,h^+ \eeq
where
\beq \Delta^N D_{h/u_V\uparrow}(z_h,Q^2)\!=\!2\,{\cal N}^{h/\!u_V}\C
(z_h)\,D_{u_V}^h(z_h,Q^2) . \eeq

 The unknown quantities are ${\cal N}^{h/\!u_V}\C (z_h)$ and $M\C  $ (often $M\C  $ is denoted
by $M$~\cite{T_1} or $M_h$~\cite{general,T_3}),  or equivalently $ \avp\C$:
\be
 \avp\C  =\frac{\avp  \, M\C ^2}{\avp  +M\C ^2}\,,
\label{Coll-frag2}
 \ee
 which characterizes the $p_\perp$-dependence. As mentioned earlier,
  $\avp$ is known from  multiplicities in unpolarized SIDIS.

%%%%%%%%%%%%%%%%%%%%%%%%%%%%%%%%%%%%%%%%%%%%%%%%%%%%%%%%%%%%%%%%%%%%%%%%%%%%%%%%%%%%%%%%%%%%%%%%%%%%%%%%%%%%%%%

%%%%%%%%%%%%%%%%%%%%%%%%%%%%%%%%%%%%%%%%%%%%%%%%%%%%%%%%%%%%%%%%%%%%%%%%%%%%%%%%%%%%%%%%%%%%%%%%%%%%%%%%%%%%%%%%%%%%%%%%%
\section{The unpolarized azimuthal and Sivers asymmetries }
%%%%%%%%%%%%%%%%%%%%%%%%%%%%%%%%%%%%%%%%%%%%%%%%%%%%%%%%%%%%%%%%%%%%%%%%%%%%%%%%%%%%%%%%%%%%%%%%%%%%%%%%%%%%%%%

The general expression for the difference cross section in SIDIS, for  unpolarized leptons on
transversely polarized nucleons, with polarization $S_T$, $l+N^\uparrow \to l +h+X$,
 in the kinematic region  $P_T \simeq k_\perp \ll Q$,   is given
in terms of the unpolarized $F_{UU}^{h-\bar h}$, $F_{UU}^{\cos \phi_h,h-\bar h}$, $F_{UU}^{\cos 2\phi_h,h-\bar h}$
 and transversely polarized  $F_{UT}^{\sin (\phi_S-\phi_h),h-\bar h}$ structure functions, by~\cite{general}:
 \be
 \frac{d\s_N^{h-\bar h}}{dx_B\,dQ^2\,dz_h\,d^2{\bf P}_T d \phi_S }&=&\frac{2\,\pi
\a^2_{em}}{Q^4}\left\{[1+(1-y)^2]\,F_{UU}^{h-\bar h}+ 2(1-y)\,\cos
2\phi_h\,F_{UU}^{\cos 2\phi_h ,h-\bar h}+\right.\nn
&&+2(2-y)\sqrt{1-y}\,\cos \phi_h\,F_{UU}^{\cos \phi_h,h-\bar h}\nn
&&\left.+ S_T\left[[1+(1-y)^2]\,\sin (\phi_S-\phi_h)\,F_{UT}^{\sin (\phi_S-\phi_h),h-\bar h}+\dots\right]\right\}\label{1}
 \ee
 Here we have kept only the terms relevant to the considerations in this paper: $ F_{UU}^{\cos 2\phi_h,h-\bar h}$ and
$F_{UU}^{\cos \phi_h ,h-\bar h}$ get contributions from both the BM
functions and the purely kinematic Cahn effect; $F_{UT}^{\sin (\phi\S-\phi_h),h-\bar h}$ gets a contribution from the Sivers function;
$F_{UU}^{h-\bar h}$ determines the unpolarized cross section without $\phi_h$-dependence.
 They involve convolutions of the corresponding valence-quark TMD parton densities and FFs~\cite{we, we0}.

Here $P_T$ is the transverse momentum of the final hadron  in the
$\g^*$-nucleon\, c.m. frame,
 and $ z_h$, $Q^2$ and $y$ are the usual measurable SIDIS quantities:
\be
 \quad z_h=\frac {(P\cdot P_h)}{(P\cdot q)},\quad Q^2=-q^2, \quad q=l-l',
\quad
 y=\frac{(P\cdot q)}{(P\cdot l)}
 \ee
 with $l$ and $l'$, $P$ and $P_h$  the 4-momenta of the initial
and final leptons,  and initial and final hadrons.  Note that
 \be
 \quad Q^2=2ME \xb y
\ee
where $M$ is the target mass (in this paper the deuteron
  mass) and $E$ the lepton laboratory energy. Throughout the
paper we
 follow the notation and kinematics of ref. \cite{general}.

In current analyses~\cite{BM_1, BM_2}, in extracting the BM function, an additional simplifying assumption is made, namely, the
BM function is taken  proportional  to
its chiral-even partner -- the Sivers function. Clearly the resulting BM function depends critically on the validity of this assumption. Our fundamental aim  is to check this key assumption using only measurable quantities --
 the difference asymmetries, and without requiring any knowledge about the TMD functions.

 The  difference azimuthal $\cos\phi_h$, $\cos 2\phi_h$ and  $\sin (\phi_S-\phi_h)$, Sivers, asymmetries that single out these terms are:
 \be
  A_{UU}^{\cos  \phi_h,h-\bar h}=\frac{\int d\phi_h\,\cos\phi_h\,d\s^{h-\bar h}}{\int d\phi_h\,d\s^{h-\bar h}}
  \ee
  \be
  A_{UU}^{\cos 2 \phi_h,h-\bar h}=\frac{\int d\phi_h\,\cos 2\phi_h\,d\s^{h-\bar h}}{\int d\phi_h\,d\s^{h-\bar h}}
  \ee
 \be
  A_{UT}^{Siv,h-\bar h}=\frac{1}{S_T}\,\frac{\int d\phi_h\,d\phi_S\,\sin (\phi_S-\phi_h)\,
  (d\s^\uparrow - d\s^\downarrow)^{h-\bar h}}{\int d\phi_h\,d\phi_S\,(d\s^\uparrow + d\s^\downarrow)^{h-\bar h}}
  \ee
The corresponding   $\xb$-dependent asymmetries, integrated over $P_T^2,\,z_h$ and $Q^2$, that we shall work with,  are:
\be
A_{UU}^{\cos \phi_h,h-\bar h}(\xb )&=&\frac{\int dQ^2\,dz_h\,dP_T^2\,[(2-y)\sqrt{1-y}/Q^4]\,F_{UU}^{\cos \phi_h,h-\bar h}}
 {\int dQ^2\,dz_h\,dP_T^2\,[[1+(1-y)^2]/Q^4]\,F_{UU}^{h-\bar h}}\label{Acos}\\
 A_{UU}^{\cos 2 \phi_h,h-\bar h}(\xb )&=&\frac{\int dQ^2\,dz_h\,dP_T^2\,[(1-y)/Q^4]\,F_{UU}^{\cos 2\phi_h,h-\bar h}}
 {\int dQ^2\,dz_h\,dP_T^2\,[[1+(1-y)^2]/Q^4]\,F_{UU}^{h-\bar h}}\label{Acos2}\\
 A_{UT}^{Siv,h-\bar h}(\xb )&=&\frac{1}{S_T}\,\frac{\int dQ^2\,dz_h\,dP_T^2\,[[1+(1-y)^2]/Q^4]\,F_{UT}^{\sin (\phi_S-\phi_h),h-\bar h}}
 {\int dQ^2\,dz_h\,dP_T^2\,[[1+(1-y)^2]/Q^4]\,F_{UU}^{h-\bar h}}
 \ee

%%%%%%%%%%%%%%%%%%%%%%%%%%%%%%%%%%%%%%%%%%%%%%%%%%%%%%%%%%%%%%%%%%%%%%%%%%%%%%%%%%%%%%%%%%%%%%%%%%%%%%%%%%%%%%
\section{Tests for the relation between the BM and Sivers functions on a deuteron target}\label{Relations 1}
%%%%%%%%%%%%%%%%%%%%%%%%%%%%%%%%%%%%%%%%%%%%%%%%%%%%%%%%%%%%%%%%%%%%%%%%%%%%%%%%%%%%%%%%%%%%%%%%%%%%%%%

In the difference asymmetries  on deuterium,
 only the sum of the valence-quarks $Q_V=u_V+d_V$ enters  for any final hadron $h$.
 Therefore, in contrast to the currently used assumption of proportionality between BM and Sivers functions for each quark and anti-quark flavour,
  we assume the simpler relation:
\be
 \Delta
f^{Q_V}\BM(x,k_\perp,Q^2)=\lambda_{Q_V}\,\Delta
f^{Q_V}_{Siv}(x,k_\perp,Q^2),\qquad Q_V=u_V+d_V\label{BM1}
 \ee
 where $\lambda_{Q_V}$ is a constant. Using  the  parametrizations (\ref{BM-Siv_dist1}),
 Eq. (\ref{BM1}) implies  that  the $k_\perp$-dependencies in BM and Sivers functions are the same,
  while the $\xb$-dependencies are proportional:
\be
M\BM=M\S,\qquad \avk\BM = \avk\S , \qquad   {\cal N}\BM^{Q_V}(\xb)=\lambda_{Q_V}\,{\cal N}^{Q_V}_{Siv}(\xb).\label{tildeBM}
\ee
The $\langle\cos \phi_h\rangle$ and $\langle\cos 2\phi_h\rangle$ azimuthal asymmetries in unpolarized SIDIS receive contributions from both the
BM function and  the purely kinematic Cahn effect.  The connection  (\ref{BM1}) between the
 BM and Sivers functions  leads to  relations between
the BM induced contributions in $\langle\cos \phi_h\rangle$ or $\langle\cos 2\phi_h\rangle$ and the Sivers asymmetries.
 Here we present the resulting relations
 between the $\xb$-dependent $\langle\cos \phi_h\rangle$ or $\langle\cos 2\phi_h\rangle$ and Sivers asymmetries.

These relations are particularly simple and predictive
if the bins in $\xb$ are small  enough, so as to neglect
the $Q^2$-evolution of the collinear functions inside the bins.

%%%%%%%%%%%%%%%%%%%%%%%%%%%%%%%%%%%%%%%%%%%%%%%%%%%%%%%%%%%%%%%%%%%%%%%%%%%%%%%%%%%%%%%%%%%%%%%%%%%%%%%%%%%%%%%%%%%%%%%%%
  \subsection{ Tests based on the asymmetry $A_{UU}^{\cos\phi_h}$}\label{Acosphi}

  Here we present the relation between the $\xb$-dependent $\cos\phi_h$  and Sivers asymmetries  on a deuteron target,
  when the  $Q^2$-evolution of the collinear parton densities and fragmentation functions
   can be neglected inside the considered $\xb$-bin.
   The standard  parametrizations (\ref{fq}), (\ref{Dq}) and  (\ref{BM-Siv_dist1}), (\ref{BM-Siv_dist2}) are used.\nl

   1. The asymmetry  $A_{UU}^{\cos\phi_h}$  has two twist-3 contributions of  $1/Q$-order from the BM function and
   from the Cahn effect. For the $\xb$-dependent difference asymmetry on a deuteron
    $A_{UU,d}^{\cos\phi_h,h-\bar h}(\xb )$ we have (see Appendix A):
   \be
   A_{UU,d}^{\cos\phi_h,h-\bar h}(\xb)=\Phi (\xb)\left\{C_{Cahn}^h+2{\cal N}\BM^{Q_V}(\xb)\,
    C\BM^h\right\},\qquad h=\pi^+, K^+, h^+\label{barAcos}
   \ee
Here  $ C_{Cahn}^h$ and $ C\BM^h$ are constants, given by:
 \be
  C_{Cahn}^h&=&-\avk\, \frac{\int dz_h\,z_h[D_{q_V}^{h}(z_h)]/\sqrt{\avPT}}{\int dz_h\, [D_{q_V}^{h}(z_h)]},\label{Cahn}\\
  C\BM^h&=&\frac{e\,\avk\BM^2\,\avp\C^2}{2\,M\BM\,M\C\avk\avp}\;\frac{\int dz_h\,[z_h^2\avk\BM  +2\avp\C]\,
 [\Delta^N D_{{q_V}\!\uparrow}^{h}(z_h)]\,
 /\avPT\BM ^{3/2}}{\int dz_h\,[D_{u_V}^{h}(z_h)]}\label{BM}\\\nn
 \avPT &=&\avp +z_h^2 \avk ,\quad \avPT\BM =\avp\C +z_h^2 \avk\BM .
 \ee
  The function $\Phi (\xb)$ is completely fixed  by kinematics, the same for all final hadrons:
\be
\Phi (\xb)=\frac{\sqrt{\pi}\,(2-\bar y)\sqrt{1-\bar y}}{\langle Q \rangle\,[1+(1-\bar y)^2]},\label{Phi}
\ee
\nl
where
\be
  \bar y = \frac{\langle Q \rangle^2}{2M_dE\,x_B}, \label{y}
\ee
$\langle Q \rangle^2$ is some mean value of $Q^2$ for each $\xb$-bin (see Appendix A), $M_d$ is the mass of the deuterium target.

 The notation $[D_{q_V}^h]$ is shorthand for the following:
\be
[D_{q_V}^{h}(z_h,Q^2)] = D_{u_V}^{h},\qquad \textrm{for} \,h=\pi^+,K^+ \label{D1}
\ee
and %while
\be
[D_{q_V}^{h^+}(z_h,Q^2)] = e_u^2\,D_{u_V}^{h^+}+e_d^2\, D_{d_V}^{h^+},  \qquad \textrm{for} \,\textrm{unidentified} \,\textrm{charged} \, \textrm{hadrons}\,\, h=h^+.
\ee
Analogously for $ [\Delta^N D_{{q_V}\!\uparrow}^{h}(z_h)]$ we have:
\be
[\Delta^N D_{{q_V}\!\uparrow}^{h}(z_h,Q^2)] &=& \Delta^N D_{{u_V}\!\uparrow}^{h}(z_h),\qquad h=\pi^+,K^+
\ee
\be
[\Delta^N D_{{q_V}\!\uparrow}^{h^+}(z_h,Q^2)] &=&
e_u^2\,\Delta^N D_{{u_V}\!\uparrow}^{h^+}+e_d^2\,\Delta^N D_{{d_V}\!\uparrow}^{h^+}.\label{D2}
\ee

 2. Following the same path, for the $\xb$-dependent Sivers difference asymmetry on a deuteron $A_{UT,d}^{Siv,h-\bar h}(\xb )$,
 when the $Q^2$-dependence in $Q_V (\xb ,Q^2)$ and in the valence-quark FFs $D_{u_V}^{h}(z_h,Q^2)$ can be neglected, we obtain~\cite{we}:
 \be
 A_{UT,d}^{Siv,h-\bar h}(\xb )&=&\frac{\sqrt{e\pi}}
 {2\sqrt2}\,A_{Siv}\, C_{Siv}^h\,{\cal N}_{Siv}^{Q_V}(\xb ),\qquad h=\pi^+, K^+, h^+\label{ASiv}\\
 A_{Siv}&=&\frac{\avk\S^2}{M\S\avk },\qquad
  C_{Siv}^h=\frac{\int dz_h\,z_h[D_{u_V}^h]/\sqrt{\avPT\S}}{\int dz_h\,[D_{u_V}^h]}\nn
 \avPT\S &=&\avp +z_h^2 \avk\S
 \ee

Note that both in $A_{UU,d}^{\cos\phi_h,h-\bar h}(\xb)$ and in $ A_{Siv,d}^{h-\bar h}(\xb )$ (a) there is no sum over quark flavour and
(b) the parton density $Q_V$ cancels out,  being the same in the numerator and denominator. \nl

 3. If the BM distribution is related to Sivers distribution by relations (\ref{BM1}) we have:
 \be
 {\cal N}\BM^{Q_V}(\xb )=\lambda_{Q_V} {\cal N}_{Siv}^{Q_V}(\xb )=\lambda_{Q_V}\,\frac {2\sqrt2}{\sqrt{e\pi}}\,\frac{1}{A_{Siv}C^h_{Siv}}\,
 A_{UT,d}^{Siv,h-\bar h}(\xb ),
 \ee
which expresses the unknown $\xb$-dependence of the BM-distribution  in terms of the measurable  $\xb$-dependent Sivers asymmetry.
The assumed relation  (\ref{BM1}) between the BM and Sivers functions then leads to the following relation between
 the $\xb$-dependent azimuthal $\cos \phi_h$-asymmetry
  $A_{UU}^{\cos \phi_h}\equiv <\cos \phi_h>$
  and the Sivers  asymmetry on a deuteron target:
\be
A_{UU,d}^{\cos\phi_h,h-\bar h}(\xb)- C_{\widetilde{BM}}^h\,\Phi (\xb)\,A_{UT,d}^{Siv,h-\bar h}(\xb )=
 C_{Cahn}^h\,\Phi (\xb), \qquad h=\pi^+, K^+, h^+ .\label{R1}
\ee
Here the function $\Phi (\xb)$ and the constant $C_{Cahn}^{h}$
are given by (\ref{Phi})
 and (\ref{Cahn}), respectively; the constant
    $C_{\widetilde{BM}}^{h}$,
 induced by the BM  function, is obtained from  the expression for the coefficient $C\BM^h$, Eq.(\ref{BM}), by making the replacements
 $M\BM \to M\S$ and $ \avk\BM \to \avk\S $, yielding:
 \beq
  C_{\widetilde{BM}}^{h}=
  2\lambda_{Q_V}\sqrt{\frac{2 e}{\pi}}\,\frac{\avp\C^2}{M\C\avp}\,
 \frac{\int_{0.2}^1 dz_h\,[z_h^2\avk\S  +2\avp\C]\,
 [\Delta^N D_{{q_V}\!\uparrow}^{h}(z_h)]\,
 /\avPT_{\widetilde{BM}}^{3/2}}{\int_{0.2}^1 dz_h\,\,z_h\,\,[D_{q_V}^{h}(z_h)]/\sqrt{\avPT\S}},
 \eeq
 where
 \be
 \avPT_{\widetilde{BM}}=\avp\C +z_h^2 \avk\S .
 \ee

  There are two important consequences of Eq. (\ref{R1}), that we shall use further:

 1) It represents a direct and simple test of the relation (\ref{BM1}) between the BM and
Sivers TMD-functions,  in which only  measurable quantities   enter,
and no knowledge about the TMD functions is required.

2) The different $\xb$-dependences  of the Cahn and BM contributions,
allow us to disentangle the Cahn contribution from the BM one in our fits to the experimental data.

%%%%%%%%%%%%%%%%%%%%%%%%%%%%%%%%%%%%%%%%%%%%%%%%%%%%%%%%%%%%%%%%%%%%%%%%%%%%%%%%%%%%%%%%%%%%%%%%%%%%%%%%%%%%%%%%%%%%%%%
 \subsection{ Tests based on  the asymmetry $A_{UU}^{\cos 2\phi_h}$}
%%%%%%%%%%%%%%%%%%%%%%%%%%%%%%%%%%%%%%%%%%%%%%%%%%%%%%%%%%%%%%%%%%%%%%%%%%%%%%%%%%%%%%%%%%%%%%%%%%%%%%%%%%%%%%%%%%%%%%%%%%%%%

 1. The asymmetry  $A_{UU}^{\cos 2\phi_h}$  has two  contributions: the leading twist-2 contribution
from BM function and the twist-4 contribution of  $1/Q^2$-order
   from the Cahn effect.

Following the same path as in obtaining Eq. (\ref{barAcos})
 (details are given in  Appendix B),  we obtain the $\xb$-dependent
difference asymmetry on a deuteron,
    $A_{UU,d}^{\cos 2\phi_h,h-\bar h}(\xb )$. The only difference is that the integration from the convolution in ${\bf k_\perp}$, in the contribution from the Cahn
    effect, cannot be carried out analytically and it
remains in the final expressions --
 these are the  integrals over $\phi$ and $k_\perp$ in Eq. (\ref{hatC}). Here we give only the final expression.

For the $\xb$-dependent difference asymmetry on a deuteron
    $A_{UU,d}^{\cos 2\phi_h,h-\bar h}(\xb )$, when the $Q^2$-dependence in $Q_V$ and in the FFs can be neglected, we obtain:
   \be
   A_{UU,d}^{\cos 2\phi_h,h-\bar h}(\xb)=\hat \Phi (\xb)\left\{{\cal N}\BM^{Q_V}(\xb)\,
   \hat C\BM^h + \frac{MM_d}{\langle Q\rangle^2}\,\hat C_{Cahn}^h\right\},\qquad h=\pi^+, K^+, h^+\label{Acos2phi}
   \ee
   where $\hat \Phi (\xb)$ is a completely fixed kinematic function, the same for all final hadrons:
\be
\hat \Phi (\xb)=\frac{2\,(1-\bar y)}{[1+(1-\bar y)^2]}.\label{hatPhi}
\ee
 The contribution from the Cahn effect is of order $1/Q^2$  compared to the BM contribution.
The constants $\hat C\BM^h$ and $\hat C_{Cahn}^h$ are:
\be
\hat C\BM^h& =& -\,\frac{e\,K}{M\BM\,M\C}\;
\frac{\int dz_h\,z_h[\Delta^ND_{q_V\uparrow }^h(z_h)]/\langle
P_T^2\rangle\BM}{\int dz_h\,[D_{q_V}^h(z_h)]},\qquad K\equiv \frac{\avk\BM^2\avp^2\C}{\avk\,\avp}
\\
\hat C_{Cahn}^{h}&=& \frac{1}{2MM_d\,\avk\avp}\; \frac{\int dz_h [D_{q_V}^{h}(z_h)]\,J(z_h)}{\int dz_h\, [D_{q_V}^{h}(z_h)]},\label{hatC}\\
 J(z_h)&\equiv &\int dP_T^2\,e^{-\frac{P_T^2}{\avp}}\int dk_\perp^2 k_\perp^2 e^{-k_\perp^2\frac{\langle P_T^2\rangle }{\avk\avp}}
 \int_0^{2\pi} d\phi\,\cos 2\phi \,e^{a\cos\phi},\quad a=\frac{2z_hk_\perp P_T}{\avp}.
 \ee

2. The  Sivers asymmetry is given in (\ref{ASiv}).\nl

3.  The assumed relation  (\ref{BM1}) between the BM and Sivers functions leads to the following relation between
 the $\xb$-dependent azimuthal $\cos 2\phi_h$-asymmetry
  $A_{UU}^{\cos 2\phi_h}\equiv \langle\cos 2\phi_h\rangle$
  and the Sivers  asymmetry on a deuteron target:
    \be
  A_{UU,d}^{\cos 2\phi_h,h-\bar h}(\xb )
  - \hat C_{\widetilde{BM}}^h\, \hat \Phi (\xb )\,  A_{UT,d}^{Siv,h-\bar h}(\xb)
  =\frac{MM_d}{\langle Q \rangle^2}\, \hat \Phi (\xb )\,\hat C_{Cahn}^h, \quad h=\pi^+, K^+, h^+.\label{R2}
   \ee

This relation and  Eq. (\ref{Acos2phi})
were previously obtained in~\cite{we} without including the $1/Q^2$-Cahn contribution.
However, as present measurements are performed at rather low $Q^2$, now we have  included  the $1/Q^2$-suppressed Cahn contribution as well.
This  is important for comparing to existing data,
which we shall do in the next Section.

 The constants $\hat C_{\widetilde{BM}}^h$ is expressed  in terms of the  parameter $\lambda_{Q_V}$
 and the TMD-fragmentation functions:
  \be
     \hat C_{\widetilde{BM}}^{h}&=& \lambda_{Q_V}\,\frac{- 2\sqrt {2e}}{\sqrt \pi}\,\frac{\avp\C^2}{M\C\avp}\,
  \frac{\int dz_h\,\,z_h\,[\Delta^N D_{{q_V}\!\uparrow}^{h}(z_h)]\,
    /\avPT_{\widetilde{BM}}} {\int
dz_h\,\,z_h\,\,[D_{q_V}^{h}(z_h)]/\sqrt{\avPT\S}},\quad h=\pi^+,K^+,h^+.
 \ee
 The FFs  $[ D_{q_V}^{h^+}(z_h)]$ and $[\Delta^N D_{{q_V}\!\uparrow}^{h^+}(z_h)]$  are given  in Eqs. (\ref{D1}) - (\ref{D2}).
%For the suggested  tests it's important that they are constants and we consider them fitted parameters.
\\

  The relations  (\ref{R1}) and (\ref{R2}) between the Sivers  $A_{UT,d}^{Siv,h-\bar h}(\xb )$
  and the unpolarized azimuthal   $ A_{UU,d}^{\cos \phi_h,h-\bar h}(\xb )$
  or $ A_{UU,d}^{\cos 2\phi_h,h-\bar h}(\xb )$ asymmetries, in which  $C_i^h$, respectively $\hat C_i^h$ are  parameters,
  represent:

   1) two independent direct tests of the assumed relation (\ref{BM1}) between the BM and
Sivers functions,  in which only  measurable quantities   enter,
and no knowledge about the TMD functions is required and,

2) two independent ways for  extracting the Cahn contribution from data.
  \nl

%%%%%%%%%%%%%%%%%%%%%%%%%%%%%%%%%%%%%%%%%%%%%%%%%%%%%%%%%%%%%%%%%%%%%%%%%%%%%%%%%%%%%%%%%%%%%%%%%%%%55
%%%%%%%%%%%%%%%%%%%%%%%%%%%%%%%%%%%%%%%%%%%%%%%%%%%%%%%%%%%%%%%%%%%%%%%%%%%%%%%5
\section{Tests using the COMPASS data for $h^\pm$ production on a deuterium target}

%%%%%%%%%%%%%%%%%%%%%%%%%%%%%%%%%%%%%%%%%%%%%%%%%%%%%%%%%%%%%%%%%%%%%%%%%%%%%%%%%%%%%%%%%%%%%%%%%%%%%%%%%%%%%%%%%%%%%
 Here we test relation (\ref{BM1}) using the COMPASS SIDIS
  data on deuteron for production of charged hadrons $h^\pm$ for the spin averaged angular distributions
  $A_{UU}^{\cos \phi_h,h^\pm}(\xb )$ and $A_{UU}^{\cos 2\phi_h,h^\pm}(\xb )$
  \cite{COMPASS-UU}, and the single-spin  Sivers asymmetry data $A_{UT}^{Siv,h^\pm}(\xb )$  \cite{COMPASS_Siv}.
   We perform the fits in three steps.

First, we   form the difference  asymmetries
 $A^{h^+-h^-}_J,\,J=\langle\cos\phi_h\rangle, \langle\cos 2\phi_h\rangle, Siv$
  from the corresponding usual asymmetries $A_j^{h^+}$ and $A_j^{h^-}$ for
positive  and negative charged hadron production \cite{COMPASS-diff}:
 \be
    A_J^{h^+-h^-} = \frac{1}{1-r} \left( A_J^{h^+} - r A_J^{h^-}\right),\qquad J=\langle\cos\phi_h\rangle, \langle\cos 2\phi_h\rangle, Siv .
    \label{eq.Diff}
\ee
Here $r$ is the ratio of the unpolarized $\xb$-dependent SIDIS
 cross sections for production of  negative and positive hadrons
 $ r = \sigma^{h^-}(\xb ) / \sigma^{h^+}(\xb )$ measured in the same kinematics \cite{COMPASS-diff}.
 As
the available data for the different  asymmetries is in  different $x_B$-bins,
  which do not match   we need to   interpolate the data.
It turns out that a linear interpolation is adequate.
Hereafter we work with the  interpolation  functions  $A_J^{h^\pm}(\xb )$ only.

 When we  determine the errors of the difference asymmetries we assume that data is not correlated.

\begin{figure}[h]
 \centerline{\includegraphics[scale=1]{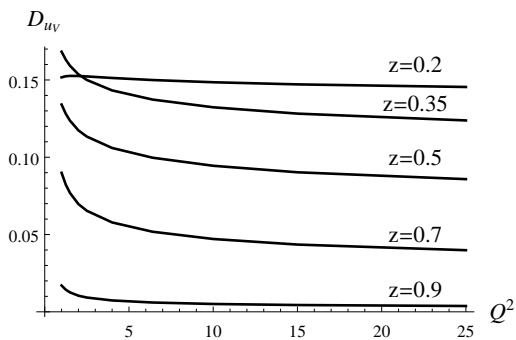}}
 \caption{The dependence of $D_{u_V}^{\pi^+}$ on $Q^2$ for different  values of $z_h=0.2,\,0.3,\,0.5,\,0.7,\,0.9$ }\label{f}
\end{figure}

Second,   we choose the $Q^2$ interval where the  $Q^2$-dependence of the collinear PDF's and FFs can be neglected.
In the COMPASS kinematics
 to each value of $\langle Q^2\rangle$ corresponds one definite value of $\langle \xb \rangle$,
  thus fixing the $Q^2$ interval we
 fix also the $\xb$-interval.
  Using the available CTEQ parametrizations for the PDFs \cite{CTEQ},  we see that there is almost no $Q^2$-dependence in
 the valence-quark distributions $u_V$ and $d_V$ in the whole
$Q^2$-range covered by COMPASS, $Q^2 \simeq [1-17]\,GeV^2$, i.e.  in the
whole $\xb$-interval.
 To get some feeling for the $Q^2$-dependence of the fragmentation function $D_{u_V}^{h^+}$ to charged hadrons,
  bearing in mind that $h^{\pm}$
  production is strongly dominated by $\pi^{\pm}$ production, in Fig. (\ref{f}) we plot
 the dependence of $D_{u_V}^{\pi^+}$ on $Q^2$  for different values of $z_h$. We use the parametrization
 in~\cite{LSS-13} obtained using the recent HERMES~\cite{HERMES-13} and preliminary COMPASS data~\cite{COMPASS-13} on multiplicities.
 This parametrization is in qualitative agreement with the one obtained from analysis of the latest COMPASS data~\cite{COMPASS-2016}.
  We see that, aside from the small values of $Q^2 \lesssim \,1.8\, GeV^2$, the $Q^2$-dependence is weak.
    We thus consider it reasonably safe to use  the following fitting interval
   $\xb\in[0.014,0.13]$  corresponding to $Q^2\in[1.77,16.27]\,GeV^2$.

Third,   we fit the parameters in Eqs~(\ref{R1}) and (\ref{R2}) using $\chi^2$-analysis. There are two ways to utilize (\ref{R1}) and (\ref{R2}), we
shall follow both of them:

($\mc{\bm{A}}$)  Provided there is enough data,   we consider  both $ C_{Cahn}^{h}$ and $ C_{\widetilde{BM}}^{h}$ (respectively
$ \hat C_{Cahn}^{h}$ and $ \hat C_{\widetilde{BM}}^{h}$)  as fitted parameters.

($\mc{\bm{B}}$) Alternatively,
 first we   calculate the Cahn constants, $ C_{Cahn}^{h}$ or $\hat C_{Cahn}^{h}$,
using the obtained expressions (\ref{Cahn}, \ref{hatC}),  and then
  fit  the same data  with just a single free parameter, $ C_{\widetilde{BM}}^{h}$
 or $ \hat C_{\widetilde{BM}}^{h}$.
 The problem with this approach, however, is that the Cahn constants depend  both on the chosen parametrizations
 for the FFs, which  don't differ so much, and on the values of the parameters
 $\avk$, $\avp$, which, as discussed in  Section (\ref{TMD-unpol}),   are rather poorly known and vary considerably.
  Consequently the main interest in this second approach will be to compare the calculated Cahn constants
  with those determined by fitting the parameters as in ($\mc{\bm{A}}$) above.

 The used $\chi^2$  for the $\langle \cos\,\phi_h\rangle^{h^\pm}$ and
 $\langle \cos\,2\phi_h\rangle^{h^\pm}$asymmetries are:
 \be
\chi^2_{cos\,\phi}&=&\int_{x_i}^{x_f} dx \,\frac{\left[F_{exp}(x)- F_{TH}(x)\right]^2}{\left[\Delta F_{exp}(x)\right]^2},\nn
\chi^2_{cos\,2\phi}&=&\int_{x_i}^{x_f} dx\, \frac{\left[\hat F_{exp}(x)-
\hat F_{TH}(x)\right]^2}{\left[\Delta \hat F_{exp}(x)\right]^2}\label{chi2}
\ee
 which take  into account the different widths of $x_B$-bins in which the data is collected.
Here $F_{exp}(\xb )$ and $\hat F_{exp}(\xb )$ denote the proper combinations of experimental data
-- the l.h.s. of eqs. (\ref{R1}) and (\ref{R2}), while $F_{TH}(\xb )$ and $\hat F_{TH}(\xb )$
are the corresponding theoretical expressions -- the r.h.s. of (\ref{R1}) and (\ref{R2}):
\be
F_{exp}(x)&=& A_{UU,d}^{\cos\phi_h,h^+-h^-}(\xb)-
C_{\widetilde{BM}}^h\,\Phi (\xb)A_{UT,d}^{Siv,h^+-h^-}(\xb ),\qquad\quad \, F_{TH}(x)=C_{Cahn}^h\,\Phi(x)\nn
\hat F_{exp}(x)&=&A_{UU,d}^{\cos 2\phi_h,h^+-h^-}(\xb )
- \hat C^h_{\widetilde{BM}}\,\hat \Phi (\xb )\,  A_{UT,d}^{Siv,h^+-h^-}(\xb),\qquad
\hat F_{TH}(x)=\frac{MM_d}{\langle Q \rangle^2(\xb)}\,\hat C_{Cahn}^h\,\hat\Phi (\xb )
\ee
 In this way the tested
relations are put in the standard form "experiment"="theory".  Note
however, that the situation here is rather peculiar because the
errors of experimental data $\Delta F_{exp}(\xb)$ and $\Delta \hat
F_{exp}(\xb)$ contain not only the
 errors of the asymmetries $\Delta
A_{UU,d}^{\cos\phi_h},\, \Delta A_{UU,d}^{\cos 2\phi_h}$ and $\Delta
A_{UU,d}^{Siv}$, but the fitting parameter as well. We have:
\be
 \Delta F_{exp}(\xb) &=&\sqrt{(\Delta
A_{UU,d}^{\cos\phi_h,h^+-h^-}(\xb))^2+
(C_{\widetilde{BM}}^h)^2\,\Phi^2(\xb)(\Delta\,A_{UT,d}^{Siv,h^+-h^-}(\xb ))^2} \label{ec}\\
\Delta \hat F_{exp}(\xb) &=&\sqrt{(\Delta A_{UU,d}^{\cos 2\phi_h,h^+-h^-}(\xb))^2+
(\hat C_{\widetilde{BM}}^h)^2\,\hat\Phi^2(\xb)(\Delta\,A_{UT,d}^{Siv,h^+-h^-}(\xb ))^2}.
\label{ec2}
 \ee
In (\ref{chi2}) the upper limit $x_f=0.13$  is fixed by the existing data for both $A_{UU,d}^{\cos\phi_h}$ and
 $A_{UU,d}^{\cos 2\phi_h}$ asymmetries,
and $x_i$ is determined  by the requirement that  it is safe to ignore  $Q^2$-variation.

\begin{figure}[h]
\begin{center}
\includegraphics[scale=1]{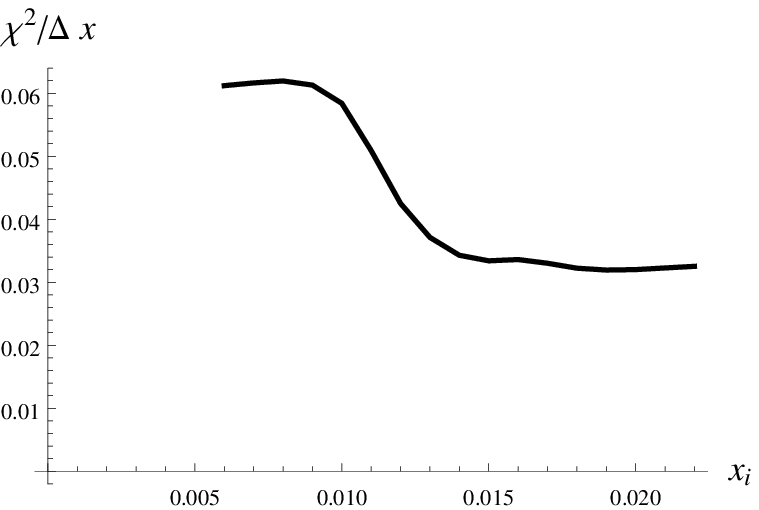}\hskip 1cm\includegraphics[scale=1]{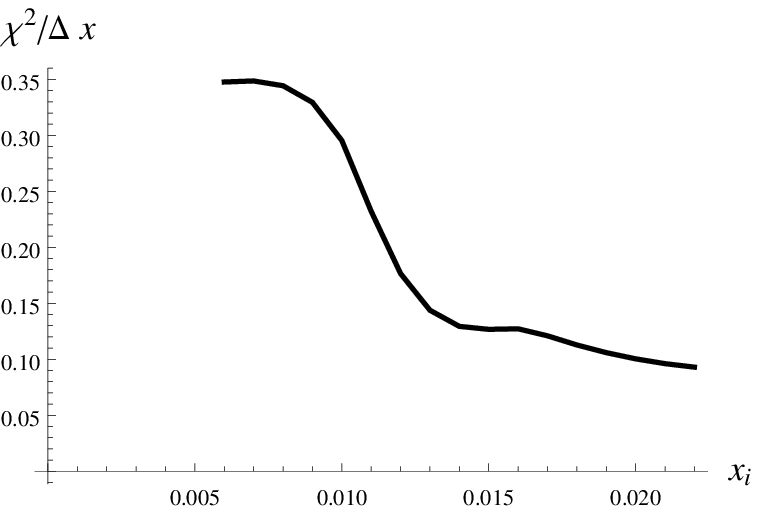}\\
a\hspace{7cm}b
\caption{The quality of the fit  $\chi^2/\Delta x$ as a function of $x_i$ with
$\Delta x = x_f -x_i$.
Panel (a) is for $\langle\cos(\phi_h)\rangle$ asymmetry, both $ C_{Cahn}$ and $C_{\widetilde{BM}}$ are fitted.
Panel (b) is for $\langle\cos(2\phi_h)\rangle$ asymmetry, both $ \hat C_{Cahn}$ and $ \hat C_{\widetilde{BM}}$ are fitted.
Note the different scales  in the two panels.}\label{cc2}
\end{center}
\end{figure}

To test quantitatively the applicability of  Eqs.~(\ref{R1}) and (\ref{R2}) for  small $\xb$ we have made series of fits
with increasing $x_i$ starting with $x_{i\,(min)}=0.006$ and going up to $x_{i\,(max)}=0.025$ and we introduce the quantity $\chi^2/\Delta x$, which is $\chi^2$ normalized to the length of the fitting interval $\Delta x = x_f -x_i$.
(It is the continuum analogue  of $\chi^2$ per  degree of freedom in the discrete case.)
The obtained  $\chi^2/\Delta x(x_i)$  functions  for both asymmetries are plotted on Fig.(\ref{cc2}).
Both of them exhibit a step-like behavior  with the step  at roughly the same position about $\xb=0.014$.
This  shows that  Eqs~(\ref{R1}, \ref{R2})  hold only for $\xb\geq 0.014$.

In the next two subsections we present the obtained values and
standard deviations of the fitted parameters. The values correspond
to the best fit of the available data with $\chi^2$ defined as
above. We use Monte Carlo
simulation in order to estimate the deviations of the fitting
parameters. On the basis of the experimental data and assuming they
have a Gaussian distribution we construct $10^3$ sets of "virtual
experimental data". For each virtual experimental data set we
determine corresponding best fit parameters.
Thus we obtain for each parameter $C^h_{\widetilde{BM}},\;\;C^h_{Cahn},\;\;\hat
C^h_{\widetilde{BM}}$ and $\hat C^h_{Cahn}$
a set with  $10^3$ data values.
Further, we filtered out the
% pairs $\{C^h_{\widetilde{BM}},\;\;C^h_{Cahn}\}$ and $\{\hat C^h_{\widetilde{BM}},\;\;\hat C^h_{Cahn}\}$
data values which are attracted by the false local minimum
corresponding not to small $\left(F_{exp}(x)- F_{TH}(x)\right)^2$ or
$\left(\hat F_{exp}(x)- \hat F_{TH}(x)\right)^2$ but to large
$\Delta F_{exp}(\xb)$ (respectively --- $\Delta \hat F_{exp}(\xb)
)$. In this way we end up with four Gaussian distributed sets for the
parameters $C^h_{\widetilde{BM}},\;\;C^h_{Cahn},\;\;\hat
C^h_{\widetilde{BM}}$ and $\hat C^h_{Cahn}$, we calculate the
standard deviation for each of them and report it as the parameter
error.

%%%%%%%%%%%%%%%%%%%%%%%%%%%%%%%%%%%%%%%%%%%%%%%%%%%%%%%%%%%%%%%%%%%%%%%%%%%%%%%%%%%%%%%%%%%%%%%%%%%%%%%%%%%%%%%%%%%%%%%%%

\subsection{ Test using the COMPASS data on $A_{UU}^{\cos\phi_h}$}\label{qq1}
%%%%%%%%%%%%%%%%%%%%%%%%%%%%%%%%%%%%%%%%%%%%%%%%%%%%%%%%%%%%%%%%%%%%%%%%%%%%%%%%%%%%%%%%%%%%%%%%%%%%%%%%%%

\begin{figure}[htb]
\begin{center}
\includegraphics[scale=1]{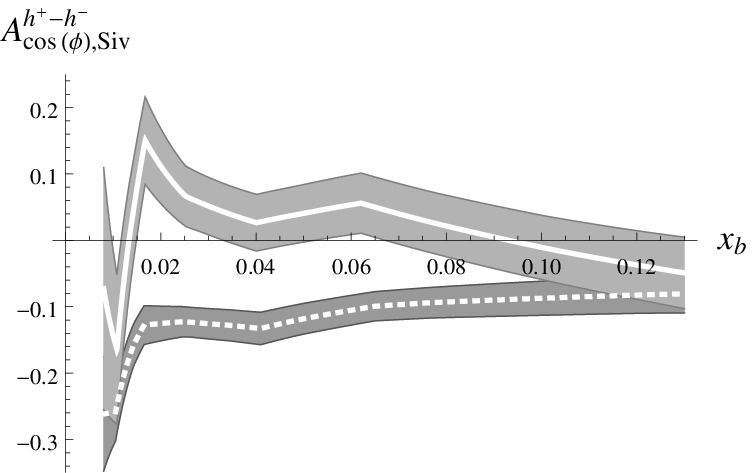}\;\;\includegraphics[scale=1]{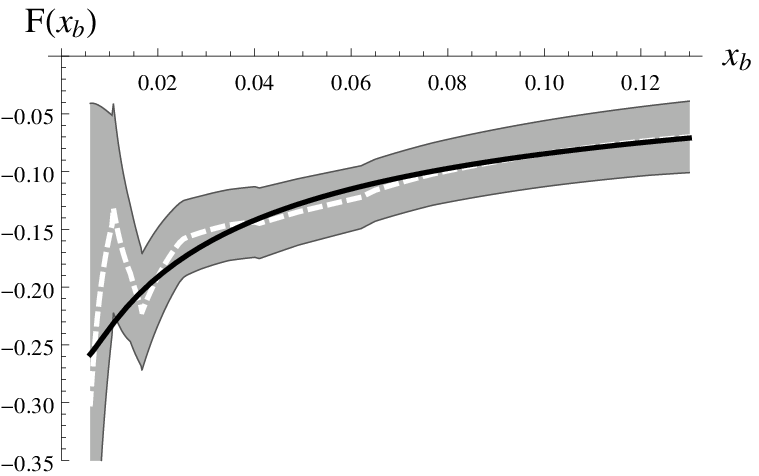}\\
a\hspace{6cm}b\\
\includegraphics[scale=1]{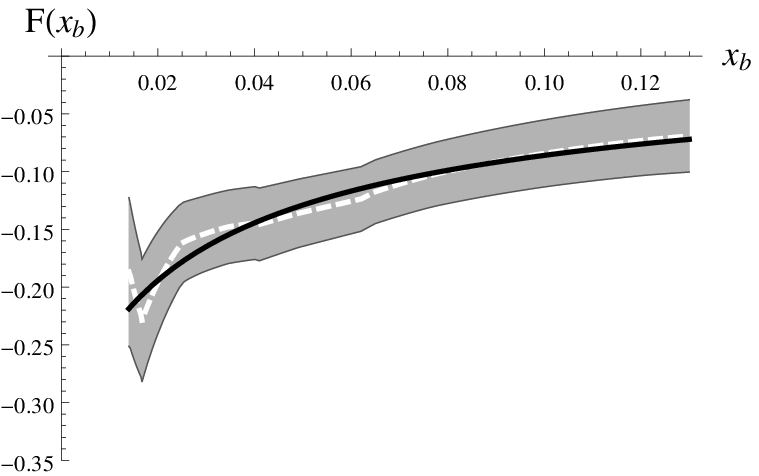}\;\;\includegraphics[scale=1]{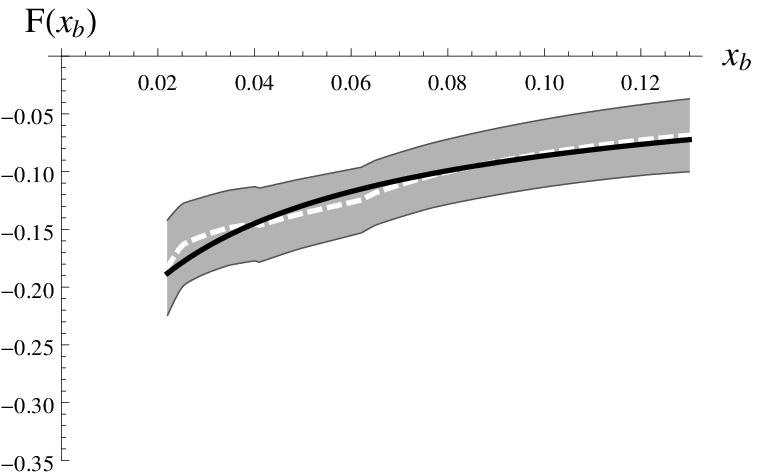}\\
c\hspace{6cm}d
\end{center}
\caption
 { The results of our tests  using $A_{UU,d}^{\cos\phi_h}(\xb)$, Eq.(\ref{R1}), following  approach ($\mc{\bm{A}}$): On panel a) are the used asymmetries $A_{UT,d}^{Siv,h^+ - h^-}(\xb )$
 (white solid line) and $A_{UU,d}^{\cos\phi_h,h^+ -h^-}(\xb)$ (white dashed line)  with their statistical errors.
On   panels  (b),  (c) and (d) are  our fits  with different $x_i$: b) $\xb \gtrsim 0.006$, c)
 $\xb \gtrsim 0.014$ and d) $\xb \gtrsim 0.022$; the dashed white lines are $F_{exp}(\xb)$ with their  errors as shaded corridors,  the solid black lines are $F_{TH}(\xb)$. } \label{fitcosphi}
\end{figure}

The difference asymmetries $A_{UU}^{\cos\phi_h, h^+-h^-}$ and $A_{UT}^{Siv, h^+-h^-}$ are presented on Fig. \ref{fitcosphi}, panel (a).
Note that the Sivers asymmetry $ A_{UT,d}^{Siv,h^+-h^-}(\xb )$ is almost zero and rather poorly determined,
which suggests,  and is proven in our fits,
 that  the corresponding fitting parameter $C^h_{\widetilde{BM}}$   will be poorly determined.

 $\bullet\, ({\cal A})$
The results of our fit in approach ($\mc{\bm{A}}$), when both $ C_{Cahn}^{h}$ and $ C_{\widetilde{BM}}^{h}$ are fitted,
 are presented on panels (b), (c) and (d) of Fig.\ref{fitcosphi}
  for the  three $\xb$-intervals:  b) $\xb \gtrsim 0.006$ ($Q^2\geq 1.15\,GeV^2$), c) for $\xb \gtrsim 0.014$
   ($Q^2\geq 1.77\,GeV^2$) and d) $\xb \gtrsim 0.022$ ($Q^2\geq 2.43\,GeV^2$).
Panels (c) and (d) show that   relation (\ref{BM1}) holds for $\xb \gtrsim 0.014$,
while the discrepancy between experiment and theory in case (b) shows that at small $\xb \lesssim 0.014$
relation (\ref{BM1}) does not hold. This agrees with  the results of Fig.\ref{cc2}a.

 $\bullet\, ({\cal B})$  In  approach ($\mc{\bm{B}}$) we need an  expression for $C_{Cahn}^h$ with integration
 over the measured $P_T$ interval in COMPASS:
\be
C_{Cahn}^h=-\frac{2}{\sqrt
\pi}\,\avk\,\frac{\int\,dz_h\,z_h\,\int
dP_T\,P_T^2\,e^{-P_T^2/\avPT}\,[D_{q_V}^h(z_h)]/\avPT^2} {\int
\,dz_h\,\int dP_T\,P_T\,e^{-P_T^2/\avPT}\,[D_{q_V}^h(z_h)]/\avPT},
\label{Cahn_calc}
\ee
 where the limits of integration are $[P_{T,min}, P_{T,max}] = [0.1, 1.0\,GeV]$ and $z_h=[0.2,\,0.85]$ \cite{COMPASS-UU}.
   (If the integration over $P_T$ is in the interval $P_T\in [0,\,\infty]$ we recover Eq. (\ref{Cahn}).)

 We need also the FF for unidentified charged hadrons $[D_{q_V}^{h^+}]$.
   To estimate this, we neglect the contribution from produced protons, (about 1\%) and  use:
 \be
 [D_{q_V}^{h^+}]\equiv e_u^2\,D_{u_V}^{h^+}+ e_d^2 \,D_{d_V}^{h^+}=
 (e_u^2-e_d^2)D_{u_V}^{\pi^+}+e_u^2 D_{u_V}^{K^+},\quad D_{q_V}^h= D_q^h-D_{\bar q}^h,
 \ee
where  we have used SU(2)-invariance for the pions,  implying:
\be
D_{u_V}^{\pi^+}=-\,D_{d_V}^{\pi^+},
\ee
and $D_{d_V}^{K^+}=0$, which follows from the quark content of kaons;
this assumption is used in all present analyses in extracting the kaon FFs.

\begin{figure}[h]
\begin{center}
\includegraphics[scale=1]{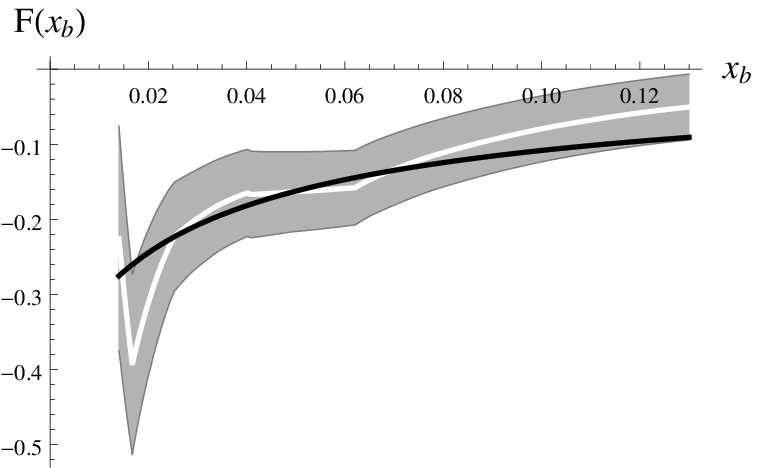}\;\;
\includegraphics[scale=1]{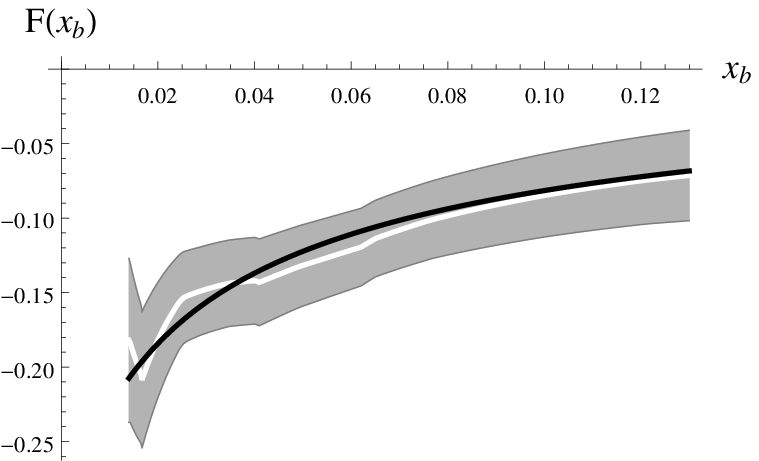}\\
a\hspace{6cm} b\\
\includegraphics[scale=1]{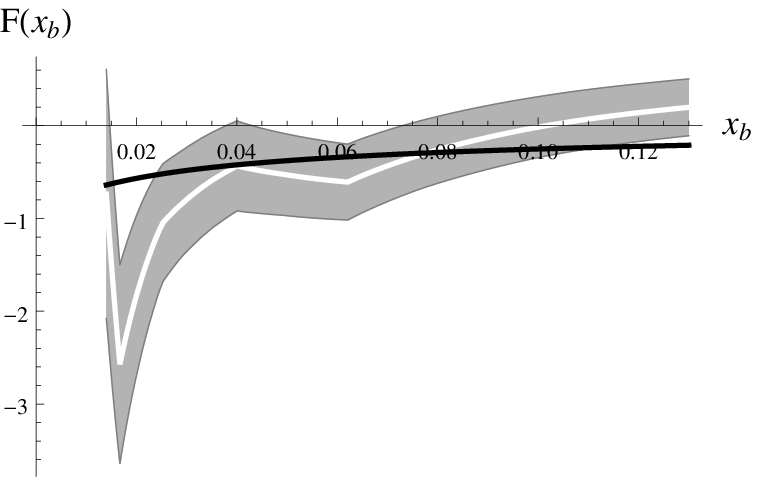}\;\;
\includegraphics[scale=1]{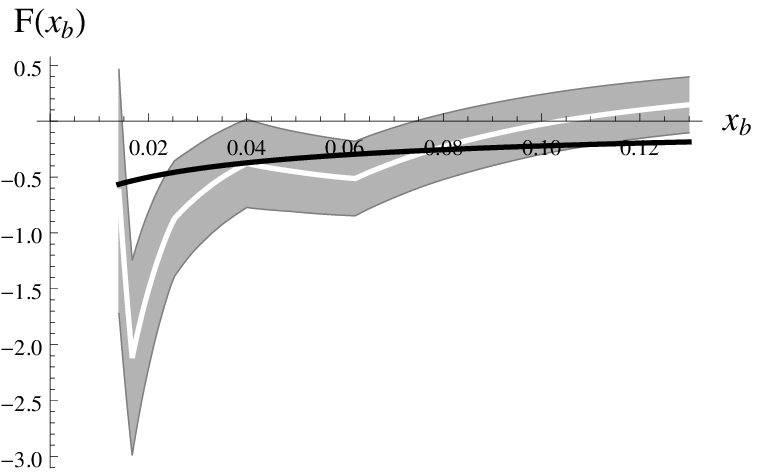}\\
c\hspace{6cm} d\\
\end{center}
\caption{The fits corresponding to approach ($\cal B$),  Table
\ref{fit2}. The white line is $F_{exp}(\xb)$ and the black one is
$F_{TH}(\xb)$ with $C_{Cahn}$ calculated with different values of
$\avk$ and $\avp$.
Panel (a) is for
$\avk=0.25\;\avp=0.20\;[\mathrm{GeV}^2]$, (b) is for
$\avk=0.18\;\avp=0.20\; [\mathrm{GeV}^2]$, (c) is for
$\avk=0.57\;\avp=0.12\; [\mathrm{GeV}^2]$ and (d) is for
$\avk=0.61\;\avp=0.19\; [\mathrm{GeV}^2]$.
 Note the different scales in the panels.} \label{cos_f_momenta}
\end{figure}

 We use two of the available parametrizations for the FFs:  AKK~\cite{AKK} and LSS~\cite{LSS-13} and find that the value of  $C_{Cahn}^h$
is not sensitive to the used parametrization; also, as expected, it is not sensitive to the chosen $\langle Q^2\rangle $.
However it is very sensitive to the values $\avk$ and $\avp$.
We find that the quality of  the fits in the approach $\mc{\bm{B}}$, with one exception,  are considerably worse than in the approach $\mc{\bm{A}}$
when both $C_{Cahn}^h$ and $C_{\widetilde{BM}}^h$ are fitted --- see Fig.\ref{cos_f_momenta}.
 The exception is for the values $\avk= 0.18,\; \avp= 0.20\; \mathrm{GeV}^2$, where the calculated $C_{Cahn}^h$
  coincides with the fitted value.
In the case of $\avk= 0.25,\; \avp= 0.20\; \mathrm{GeV}^2$ the calculated $C_{Cahn}^h$ is within the error of the fitted one.

 For $\avk= 0.57,\; \avp= 0.12\; \mathrm{GeV}^2$ the discrepancy between calculated and fitted  $C_{Cahn}^h$ is $4.9 \;\sigma$ and goes up to $5.8 \;\sigma$ for $C_{\widetilde{BM}}^h$.
 For $\avk= 0.61,\; \avp= 0.19\; \mathrm{GeV}^2$ the discrepancy is $2.1 \;\sigma$ for $C_{Cahn}^h$ and $2.6 \;\sigma$ for $C_{\widetilde{BM}}^h$.

  This can be verified also in Table \ref{fit2},  where the obtained  numerical values  for  $C_{Cahn}^h$ and $C^h_{\widetilde{BM}}$
 in approaches ($\cal A$) and ($\cal B$)
  are presented. The
presented errors correspond to 1  standard deviation.
Note that from the analytic expression   Eq.~(\ref{Cahn}), it follows that   $C_{Cahn}^h$ should be negative,
which is  in agreement with the value obtained from the fit.

\begin{table}[h]
\begin{tabular}{|c|c|c|c|c|c|}
  \hline
&$\mc{\bm{A}}$& \multicolumn{4}{c|}{$\mc{\bm{B}}$}\\
\hline
$\avk\; [\mathrm{GeV}^2]$ &&
 0.25 & 0.18 & $ 0.57 \pm 0.08$ & $0.61 \pm 0.20$\\
$\avp\; [\mathrm{GeV}^2]$ &&
 0.20 & 0.20 & $0.12 \pm 0.01$ & $0.19 \pm 0.02$\\
\hline
  $C_{Cahn}^h$&$-0.167\pm 0.043$ &
 -0.21&
 -0.16 &
 $-0.49 \pm 0.05$ &
 $-0.43 \pm 0.10$\\
  \hline
$C^h_{\widetilde{BM}}$
   & $0.55\pm 0.80$ & $1.43\;(\pm 1.7)$& $0.44\;(\pm 0.93)$ & $13 \pm 2\; (\pm 6)$ &
   $11 \pm 4\; (\pm 6) $    \\
 \hline
$ \chi^2/\Delta x$
   & 0.034 & 0.27 & 0.055 & 0.58 & 0.57\\
 \hline
\end{tabular}
\caption{ The  numerical values for the parameters: ($\mc{\bm{A}}$): Both $C_{Cahn}^h$ and  $C_{\widetilde{BM}}^h$  are obtained fitting Eq.(\ref{R1}),
the errors (which correspond to 1 standard
deviation) are obtained with MC simulation.
($\mc{\bm{B}}$): $C_{Cahn}^h$ is calculated using Eq.~(\ref{Cahn_calc}) in which the FFs are from  LSS~\cite{LSS-13}, and the values for $\avk$ and $\avp$  are those discussed in Sec.\ref{TMD-unpol}.
 $C_{\widetilde{BM}}^h$  is obtained fitting Eq.(\ref{R1}).
  We give two errors.
 First we give the error due to momenta spread (only given for the cases where errors of  $\avk$, $\avp$ are known).
 Second, in parentheses, we give the total standard deviation due to both momenta and data errors. }
\label{fit2}
\end{table}

To the best of our knowledge this is the first time that the Cahn contribution $C_{Cahn}^h$ has been determined from data and it
is puzzling that its value is in agreement with a calculated result based on  the early values of the Gaussian parameters
$\avk= 0.18,\; \avp= 0.20\; \mathrm{GeV}^2$, which are supposed to be  ruled out by later measurements.

%%%%%%%%%%%%%%%%%%%%%%%%%%%%%%%%%%%%%%%%%%%%%%%%%%%%%%%%%%%%%%%%%%%%%%%%%%%%%%%%%%%%%%%%%%%%%%%%%%%%%%%%%
%%%%%%%%%%%%%%%%%%%%%%%%%%%%%%%%%%%%%%%%%%%%%%%%%%%%%%%%%%%%%%%%%%%%%%%%%%%%%%%%%%%%%%%%%%%%%%%%%%%%%%%
\subsection{Test using the COMPASS data on $A_{UU}^{\cos 2\phi_h}$}\label{qq2}

\begin{figure}[h]
\begin{center}
\includegraphics[scale=1]{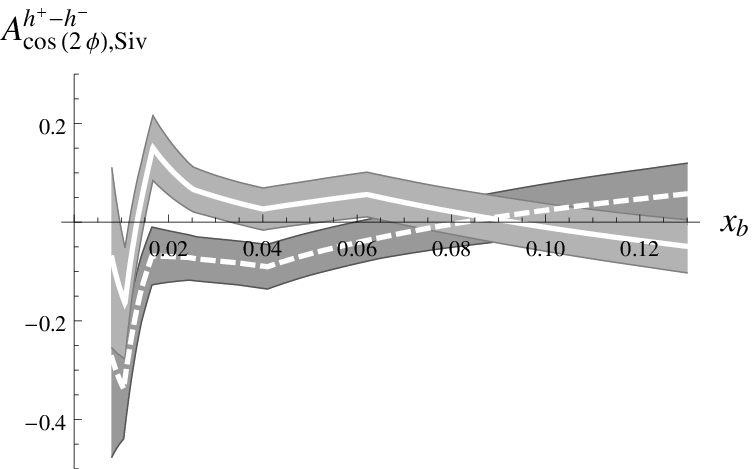}\;\;
\includegraphics[scale=1]{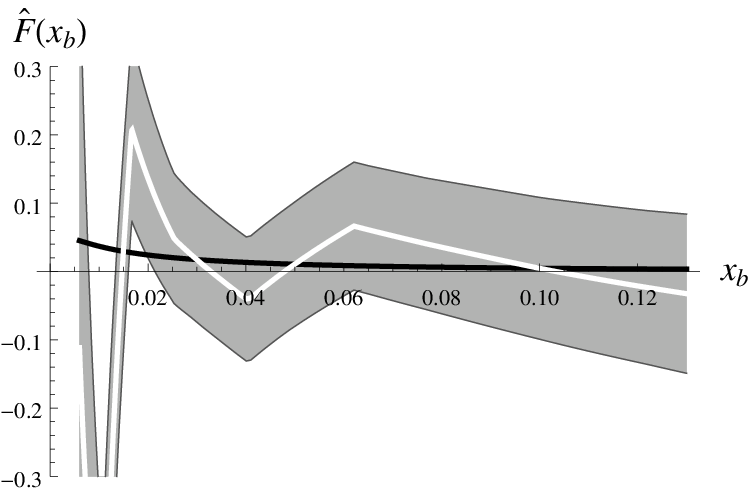}\\
a\hspace{6cm}b\\
\includegraphics[scale=1]{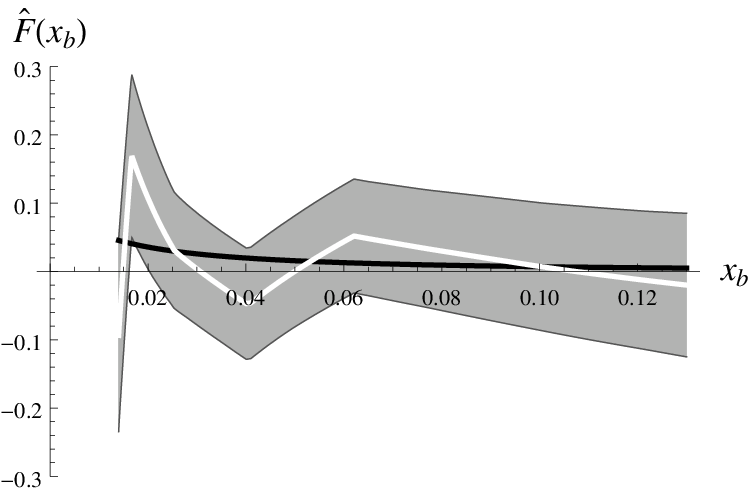}\;\;
\includegraphics[scale=1]{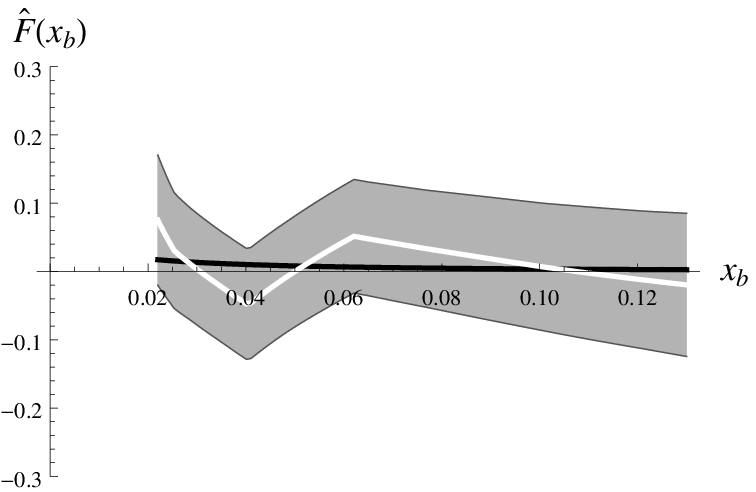}\\
c\hspace{6cm}d
\end{center}
\caption{  The results of our tests using $A_{UU,d}^{\cos2\phi_h}(\xb)$,  Eq.(\ref{R2}), following  approach ($\mc{\bm{A}}$): On panel a) are the used asymmetries $A_{UT,d}^{Siv,h^+ - h^-}(\xb )$
 (white solid line) and $A_{UU,d}^{\cos 2\phi_h,h^+ -h^-}(\xb)$ (white dashed line)  with their statistical errors.
On   panels  (b),  (c) and (d) are  our fits for different $x_i$: b) $\xb \gtrsim 0.006$, c)
 $\xb \gtrsim 0.014$ and d) $\xb \gtrsim 0.022$; the  white lines are $\hat F_{exp}(\xb)$ with their  errors as shaded corridors,  the black lines are $\hat F_{TH}(\xb)$.}
\label{fitcos2phi}
\end{figure}

 The used difference asymmetries $A_{UU}^{\cos 2\phi_h, h^+-h^-}$ and $A_{UT}^{Siv, h^+-h^-}$ are presented in Fig. \ref{fitcos2phi}a.
Note that now both  asymmetries  are poorly determined with large relative errors,
which implies that both fitting parameters  ($\hat C^h_{\widetilde{BM}}$ and $\hat C^h_{Cahn}$) will be poorly determined.
In Fig. \ref{fitcos2phi} we show the fit to Eq. (\ref{R2}) in approach ($\cal A$): panel b) is the fit for $\xb \gtrsim 0.006$. The interval of discrepancy  between experiment
 and theory is clearly visible. Panels c) and d) are for fits corresponding to the kinematics of the right "plateau" of the $\chi^2$ function (Fig. \ref{cc2}b).
 Panels c)  and d) are  for $\xb \gtrsim 0.014$ and $\xb \gtrsim 0.022$, respectively.
Analogously as for the $\langle \cos\, \phi_h\rangle^{h^\pm}$-case, the theoretical function is within the experimental margins
 for $x_i \gtrsim 0.014$, however the relative errors in the present case are considerably bigger.
   Note that in the range $\xb >0.014$, $A^{\cos 2\phi_h,h^+-h^-}_{UU,d}$ and $A^{Siv,h^+-h^-}_{UT,d}$ have opposite signs,
     which suggests a small contribution from the Cahn effect. This  follows also from our theoretical formula Eq. (\ref{R2})
     and is confirmed  by the obtained numerical values for $\hat{C}_{Cahn}^h$ and $\hat{C}^h_{\widetilde{BM}}$  summarized in Table \ref{fit4}.

The results of approaches ($\mc{\bm{A}}$) and ($\mc{\bm{B}}$) are compared in Table \ref{fit4}.
The
errors cited therein correspond to 1 standard deviation. As expected, $\hat{C}^h_{\widetilde {BM}}$ is very poorly determined.

 Here the calculated and fitted values of  $\hat{C}_{Cahn}^h$ agree for $\avk= 0.25,\; \avp= 0.20$ and  $\avk= 0.18,\; \avp= 0.20\;\mathrm{GeV}^2$ --- a result similar to the one found from the $\langle\cos\phi_h\rangle$ asymmetry.
 For $\avk= 0.57,\; \avp= 0.12\; \mathrm{GeV}^2$ the discrepancy between calculated and fitted  $\hat{C}_{Cahn}^h$ is $1.4 \;\sigma$ and it goes up to $1.8 \;\sigma$ for $\hat{C}^h_{\widetilde{BM}}$.
 For $\avk= 0.61,\; \avp= 0.19\; \mathrm{GeV}^2$ the discrepancy is $0.9 \;\sigma$ for $\hat{C}_{Cahn}^h$ and $1.1 \;\sigma$ for $\hat{C}^h_{\widetilde{BM}}$.

\begin{table}[h]
\begin{tabular}{|c|c|c|c|c|c|}
  \hline
&$\mc{\bm{A}}$& \multicolumn{4}{c|}{$\mc{\bm{B}}$}\\
\hline
$\avk\; [\mathrm{GeV}^2]$ &&
 0.25 & 0.18 & $ 0.57 \pm 0.08$ & $0.61 \pm 0.20$\\
$\avp\; [\mathrm{GeV}^2]$ &&
 0.20 & 0.20 & $0.12 \pm 0.01$ & $0.19 \pm 0.02$\\
\hline
  $\hat{C}_{Cahn}^h$& $0.083\pm 0.22$&
 0.079 &
 0.045 &
$ 0.41 \pm 0.08 $&
$ 0.32 \pm 0.15$\\
  \hline
$\hat{C}^h_{\widetilde{BM}}$
   &$-1.6\pm 1.6$ & $-1.9 \;(\pm 1.9)$ & $-1.6\;(\pm 1.4)$ & $-4.9 \pm 0.8\;(\pm 3.0)$& $-4.0 \pm 1.5\; (\pm 3.0) $  \\
 \hline
$ \chi^2/\Delta x$
   & 0.13 & 0.15 & 0.16 & 0.20 & 0.19\\
 \hline
\end{tabular}
\caption{The  numerical values for the parameters:
 ($\mc{\bm{A}}$): Both $\hat{C}_{Cahn}^h$ and  $\hat{C}_{\widetilde{BM}}^h$  are obtained fitting Eq.(\ref{R2}),
the errors (which correspond to 1 standard
deviation) are obtained with MC simulation.
($\mc{\bm{B}}$): $\hat{C}_{Cahn}^h$ is calculated using Eq.~(\ref{hatC}) in which the FFs are from  LSS~\cite{LSS-13},
and the values for $\avk$ and $\avp$  are those  discussed in Sec.\ref{TMD-unpol}.
 $\hat{C}_{\widetilde{BM}}^h$  is obtained  fitting Eq.(\ref{R2}).
 We give two errors.
 First we give the error due to momenta spread (only given for the cases where errors of  $\avk$, $\avp$ are known).
 Second, in parentheses, we give the total standard deviation due to both momenta and data errors.}
\label{fit4}
\end{table}

%%%%%%%%%%%%%%%%%%%%%%%%%%%%%%%%%%%%%%%%%%%%%%%%%%%%%%%%%%%%%%%%%%%%%%%%%%%%%%%%%%%%%%%%%%%%%%%%%%%%%%%%%%%%%%%
\subsection{Comparision to the existing published extraction of the BM functions  \cite{BM_1,BM_2}}\label{qc5}

In this paper we have tested the assumption of proportionality of the BM and
Sivers functions  for the sum of valence quarks $Q_V=u_V+d_V$,
(eq.\ref{BM1}). However, in ref.\cite{BM_1,BM_2} the BM
functions have been extracted from the $\cos 2 \phi$ asymmetry
assuming proportionality for each quark and anti-quark flavor $q$ separately:
\beq
\Delta f^q \BM (\xb ,\kt )= \lambda_q \Delta f^q_{Siv}(\xb ,\kt )
\label{te}
 \eeq
A legitimate  question arises as to the compatibility of the two approaches  i.e whether Eqs.(\ref{te}) and (\ref{BM1})  are compatible. Here we study this question.

Under the assumption of eqs.(\ref{te}) one obtains:
\beq
4\,\Delta f^{Q_V}\BM = (\lambda_u + \lambda_d + \lambda_{\bar u} + \lambda_{\bar d})\Delta  f^{Q_V}_{Siv} + \Delta \label{they}
\eeq
where
\be
\Delta &=& (3\lambda_u - \lambda_d - \lambda_{\bar u} - \lambda_{\bar d})\,\Delta  f_{Siv}^u
 +(\lambda_u + \lambda_d -3 \lambda_{\bar u} + \lambda_{\bar d})\, \Delta f_{Siv}^{\bar u} \nn
&&+(-\lambda_u + 3\lambda_d - \lambda_{\bar u} - \lambda_{\bar d})\, f_{Siv}^d
+(\lambda_u + \lambda_d + \lambda_{\bar u} -3 \lambda_{\bar d})\, \Delta  f_{Siv}^{\bar d}
\ee
Eq. (\ref{they}) is compatible with our assumption of proportionality Eq.(\ref{BM1})  if:
\beq
\vert \Delta \vert  \ll \vert (\lambda_u + \lambda_d + \lambda_{\bar u} + \lambda_{\bar d}) \Delta f^{Q_V}_{Siv} \vert\cdot
\label{compare}
\eeq
Note that at
\be
\lambda_u = \lambda_d = \lambda_{\bar u} = \lambda_{\bar d}\equiv \lambda_{Q_V}
\ee
we have $\Delta =0$ and we obtain Eq.(\ref{BM1}).

\begin{figure}[h]
\begin{center}
\includegraphics[scale=1]{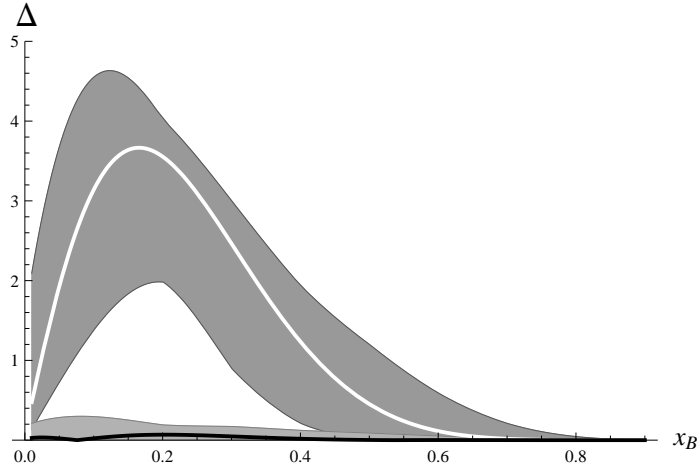}
\end{center}
\caption{A comparison between  $\vert(\lambda_u + \lambda_d + \lambda_{\bar u} +
\lambda_{\bar d})\Delta  f^{Q_V}\S\vert$ (the black curve which is almost $0$) and $\vert\Delta\vert$ (white curve).
The shaded areas are the corresponding statistical errors. The parametrization of $ f^q_{Siv}$ is taken from \cite{Anselmino_Siv}
 and the values of $\lambda_q$ are from \cite{BM_2}.
}
\label{fgf}
\end{figure}

The values for $\lambda_{u,\;d}$  are those obtained in \cite{BM_2} assuming
$
\lambda_{\bar u}= -1,\; \lambda_{\bar d} = +1 $ for the antiquarks i.e.
\be
\lambda_u=2.1\pm 0.1,\qquad \lambda_d=-1.111\pm 0.001.
\ee

The parametrization of the Sivers function for each quark flavour is taken from \cite{Anselmino_Siv}:
\beq
\hspace*{-.5cm}  \Delta  f^{q}_{Siv} (\xb,\kt ) \!=\! \Delta  f^{q}_{Siv} (\xb )\; \sqrt{2e}\,\frac{\kt}{M_1} \;
\frac{e^{-\kt^2/\avk\S }}{\pi\avk\S }
\eeq
with
\beq
\Delta
f^{q}_{Siv}(\xb )\!=\! 2\,{\cal N}\S^{q}(\xb)\,q(\xb ) \label{BM-Siv_dist3}
 \eeq
\be
{\cal N}_{q}(x) = N_q\, x^{\a_q} (1 - x) ^{\b_q}\frac{(\a_q +
\b_q)^{\a_q + \b_q}}{\a_q^{\a_q}\b_q^{\b_q}} \ee where: \be N_u &= &
0.35\pm 0.08,\qquad N_d = -0.90^{+0.43}_{-0.10}\nn N_{\bar u} &=&
0.04^{+0.22}_{-0.24},\qquad  N_{\bar d} = -0.40^{+0.33}_{-0.44}\nn
\a_u &=& 0.73^{+0.72}_{-0.58},\qquad \a_d = 1.08^{+0.82}_{-0.65},
\qquad \a_{sea} = 0.79^{+0.56}_{-0.47}\nn \b_q &\equiv &
\b=3.46^{+4.87}_{-2.90},\qquad
 M_1^2 = 0.34^{+0.30}_{-0.16}\, (GeV/c)^2
\ee
As the  dependence on $k_\perp^2$ is the same for both the BM and Sivers functions, in  Fig.(\ref{fgf}) we compare only
the dependence on $\xb$ of the two functions
 $(\lambda_u + \lambda_d + \lambda_{\bar u} + \lambda_{\bar d})\Delta  f^{Q_V}_{Siv} (\xb ) $ and $ \Delta (\xb )$.
  For the unpolarized PDFs the CTEQ6 parametrization was used.

 From this figure it
 is clear that, even accounting for the enormous errors induced by the errors of the Sivers functons,
  $\vert \Delta \vert$ is {\it much  bigger} than
 $\vert (\lambda_u + \lambda_d + \lambda_{\bar u} + \lambda_{\bar d}) \Delta f^{Q_V}_{Siv} \vert$, which is just the opposite to
 Eq. (\ref{compare}). This suggests that the extraction of the Boer-Mulders function in the literature \cite{BM_1,BM_2} is unreliable.

%%%%%%%%%%%%%%%%%%%%%%%%%%%%%%%%%%%%%%%%%%%%%%%%%%%%%%%%%%%%%%%%%%%%%%%%%%%%%%%%%%%%%%%%%%%%%%%%%%%%%%%%%%%%%%%

%%%%%%%%%%%%%%%%%%%%%%%%%%%%%%%%%%%%%%%%%%%%%%%%%%%%%%%%%%%%%%%%%%%%%%%%%%%%%%%%%%%%%%%%%%%
\section{Conclusions}
%%%%%%%%%%%%%%%%%%%%%%%%%%%%%%%%%%%%%%%%%%%%%%%%%%%%%%%%%%%%%%%%%%%%%%%%%%%%%%%%%%%%%%%%%%%%%

We had shown  previously \cite{we} that data on difference asymmetries allow one to test the assumed relation
 of proportionality between the BM and Sivers functions, which is currently used in
 the extraction of the BM function from data.
In the present paper  we  perform two independent tests of this assumption applied, however,
 to the sum of the valence-quark TMD distributions,  (\ref{BM1}), using the COMPASS SIDIS data \cite{COMPASS-UU,COMPASS_Siv} on the
 difference asymmetries
$A_{UU,d}^{\cos\phi_h,h^+ -h^-}(\xb)$,  $A_{UU,d}^{\cos 2\phi_h,h^+ -h^-}(\xb)$
 and $A_{UT,d}^{Siv,h^+ - h^-}(\xb )$. Both tests are consistent with this  assumption
  in the same  kinematic interval $\xb = [0.014 , 0.13]$.

 However, in the published extractions of the  BM functions  \cite{BM_1,BM_2},  obtained in a completely different kind of analysis, based on the available parametrizations of both Sivers and Collins functions, it is assumed that
BM and Sivers functions are proportional for each quark and anti-quark separately (Eq.(\ref{te})).
 This  would agree with our result, based  only on measurable quantities, if
$\lambda_u \approx \lambda_{\bar u}\approx\lambda_d \approx\lambda_{\bar d}\approx\lambda_{Q_V}$, which does not correspond to the values and their errors obtained in \cite{BM_1,BM_2}.

We have also determined the kinematical Cahn contribution, both directly from a fit to the data  (as far as we know for the first time) and from a calculation.
  The calculated values are very sensitive to  the average transverse momentum-squared, $\avk$ and $\avp$
 in the unpolarized PDFs and FFs, respectively. Surprisingly,
  both for $A_{UU,d}^{\cos \phi_h,h^+ -h^-}(\xb)$ and  $A_{UU,d}^{\cos 2\phi_h,h^+ -h^-}(\xb)$, the calculated values
 agree with the extracted ones
only for average transverse momenta close to the old experimental values,
 $\avk = 0.18  $
 and $\avk = 0.25,\,  \avp = 0.20 \,\mathrm{GeV}^2$  and completely disagree with the much bigger present-day estimates.
 On smaller values for the intrinsic  transverse momenta was suggested also in the covariant parton model~\cite{Zavada,Elliot}.
\\

%%%%%%%%%%%%%%%%%%%%%%%%%%%%%%%%%%%%%%%%%%%%%%%%%%%%%%%%%%%%%%%%%%%%%%%%%%%%%%%%%%%%%%%%%%%%%%%%%%%%%%%%%%%%%%%%%%%%%%%%%%%%%%%%%%%%%%%%%
\acknowledgments
%%%%%%%%%%%%%%%%%%%%%%%%%%%%%%%%%%%%%%%%%%%%%%%%%%%%%%%%%%%%%%%%%%%%%%%%%%%%%%%%%%%%%%%%%%%%%%%%%%%%%%%%%%%%%%%%%%%%%%%%%%%%%%%%%%%%%%%%%
%%%%%%%%%%%%%%%%%%%%%%%%%%%%%%%%%%%%%%%%%%%%%%%%%%%%%%%%%%%%%%%%%%%%%%%%%%%%%%%%%%%%%%%%%%%%%%%%%%%%%%%%%%%%%%%%%%%%%%%%%%%%%%%%%%%%%%%%%

 We are grateful to Fabienne Kunne and Anna Martin for helpful comments
concerning the  COMPASS data, and  E.Ch. acknowledges helpful
discussions on the Cahn effect.
 E.Ch. and M.S. acknowledge the support of Grant 08-17/2016  of the Bulgarian Science Foundation.
 E. L. is grateful to the  Leverhulme Trust for an Emeritus Fellowship.

%%%%%%%%%%%%%%%%%%%%%%%%%%%%%%%%%%%%%%%%%%%%%%%%%%%%%%%%%%%%%%%%%%%%%%%%%%%%%%%%%%%%%%%%%%%%%%%%%%%%%%%%%%%%
\section*{Appendix A: The $A_{UU}^{\cos\phi_h}$ asymmetry}

 The structure function  $F_{UU}^{\cos\phi_h}$ that determines the azimuthal $A_{UU}^{\cos\phi_h}$ asymmetry, Eq. (\ref{Acos}),
 has two twist-3 contributions of  $1/Q$-order from the Cahn effect and the BM TMDs:
\be
F_{UU}^{\cos\phi_h} = F_{Cahn}^{\cos\phi_h}+F_{BM}^{\cos\phi_h}
\ee
For the difference cross sections  $(h-\bar h)$ on deuteron
target  it is
 only the sum of the valence-quark parton densities $Q_V$ enter these functions and for $h=\pi^+,K^+,h^+$ they read~\cite{we0}:
\be
F_{Cahn,d}^{\cos\phi_h,h-\bar h}&=&\frac{2}{Q}\,Q_V(\xb ,Q^2 ) [D_{q_V}^h (z_h,Q^2)]\,{\cal A}_{Cahn}^{\cos\phi_h}(z_h,P_T^2)\\
F_{BM,d}^{\cos\phi_h,h-\bar h}&=&\frac{2}{Q}\,\Delta f\BM^{Q_V}(\xb
,Q^2)
 [\Delta^N D^h_{q_V\uparrow } (z_h,Q^2)]\,{\cal A}\BM^{\cos\phi_h}(z_h,P_T^2)
\ee
The functions  ${\cal A}_{Cahn}^{\cos\phi_h}$ and ${\cal
A}\BM^{\cos\phi_h}$ are independent of quark flavour and of the
final hadron $h$:
\be
{\cal A}_{Cahn}^{\cos\phi_h}(z_h, P_T^2)&=&-P_T z_h\,\avk \,\frac{e^{-P_T^2/\avPT }}{\pi \avPT^2}\\
{\cal A}\BM^{\cos\phi_h} (z_h, P_T^2)&=&e\,K\,P_T
\frac{e^{-P_T^2/\avPT\BM }}{\pi \avPT^4\BM}\,
\left[\left(-z_h^2\avk\BM+\avp\C\right)
\avPT\BM+z_h^2\avk\BM\,P_T^2\right]\nn
 K&\equiv &\frac{\avk\BM^2\avp^2\C}{\avk\,\avp}\label{K}
\ee

Here the notation $[D_{q_V}^h]$ and $[\Delta^N D_{{q_V}\!\uparrow}^{h}(z_h)]$ stand for combinations of the valence-quark
collinear and TMD FFs defined in Eqs. (\ref{D1})-(\ref{D2}).

We can perform the  integration over $P_T$ analytically and we
obtain:
\be
\int dP_T^2\,F_{UU,d}^{\cos\phi_h,h-\bar h}&=&
 \frac{2}{\sqrt \pi Q}\,Q_V(\xb ,Q^2)\left\{A^h_{Cahn}(z_h.Q^2)+2{\cal N}\BM(\xb)\,A^h\BM (z_h,Q^2)\right\}
\ee
where we have used the standard parametrization Eq.
(\ref{BM-Siv_dist1}) for the BM function, the notation $A^h_{Cahn,BM}$
stands for:
 \be
A^h_{Cahn}(z_h.Q^2)&\equiv &\sqrt{\pi}\,\,[D_{u_V}^h(z_h,Q^2)]\int
dP_T^2\,{\cal A}_{Cahn}^{\cos\phi_h}(z_h, P_T^2)
=-\,\frac{z_h \avk}{2\sqrt{\avPT}}\,[D_{u_V}^h(z_h,Q^2)]\\
A^h\BM (z_h.Q^2)&\equiv &\sqrt{\pi}\,\,[\Delta^N D_{u_V\uparrow
}^h(z_h,Q^2)]\int dP_T^2\,{\cal A}\BM^{\cos\phi_h}(z_h, P_T^2)\nn
&=&\frac{e\,K}{4M\BM
M\C\avPT\BM^{3/2}}\,\left[z_h^2\avk\BM+2\avp\C\right]
\left[\Delta^ND_{u_V\uparrow}^h(z_h,Q^2)\right].
\ee

For the unpolarized function $F_{UU,d}^{h-\bar h}$, that normalizes
the asymmetry, we have:
\be
\int dP_T^2\,F_{UU,d}^{h-\bar
h}=\frac{1}{\pi}\,Q_V(\xb ,Q^2)\,[D_{u_V}^h(z_h,Q^2)]\label{UU}
\ee

Thus, for the integrated over $P_T^2$ asymmetry
$A_{UU,d}^{\cos\phi_h,h-\bar h}$ we obtain:
 \be
A_{UU,d}^{\cos\phi_h,h-\bar h} (\xb ,z_h,Q^2)=\frac{2\sqrt\pi
[(2-y)\sqrt {1-y}/Q^5]\,Q_V(\xb ,Q^2)
\left[A^h_{Cahn}(z_h.Q^2)+2{\cal N}\BM(\xb)\,A^h\BM (z_h,Q^2)\right
]}{[[1+(1-y)^2]/Q^4]\,Q_V(\xb ,Q^2)\,[D_{u_V}^h(z_h,Q^2)]}
 \ee

>From this expression it follows, that if one can neglect
$Q^2$-dependence in $Q_V (\xb ,Q^2)$ and in the FFs, the $\xb$- and
$z_h$-dependencies will factorize. Also,  $Q_V (\xb)$ in the numerator
and denominator   cancel out and for the $\xb$-dependent difference
asymmetry on deuteron   $A_{UU,d}^{\cos\phi_h,h-\bar h}(\xb )$ we
obtain:
 \be
 A_{UU,d}^{\cos\phi_h,h-\bar h} (\xb )= \frac{2\sqrt \pi \int
dQ^2\,[(2-y)\sqrt {1-y}/Q^5]\, \,\int d
z_h\,\left[A^h_{Cahn}(z_h,Q^2)+2{\cal N}\BM(\xb)\,A^h\BM (z_h,Q^2)\right ]} {\int d Q^2\,[1+(1-y)^2]/Q^4]\,\int d
z_h\,[D_{u_V}^h(z_h,Q^2)]}\label{Axb}
\ee
 Further, after neglecting $Q^2$-dependence in the collinear FFs, and replacing the integration over $Q^2$ by $\Delta Q^2$ times
the function evaluated at some average value $\langle Q\rangle$ (or equivalently $\bar y$) for each $\xb$-bin,
 we obtain the simple $\xb$-dependent expression for the asymmetry:
   \be
   A_{UU,d}^{\cos\phi_h,h-\bar h}(\xb)=\Phi (\xb)\left\{C_{Cahn}^h+2{\cal N}\BM^{Q_V}(\xb)\,
    C\BM^h\right\},\qquad h=\pi^+, K^+, h^+
   \ee
    The function $\Phi (\xb)$ is given in Eq. (\ref{Phi}), it is  completely fixed  by kinematics, the same for all final hadrons.
The constants $ C_{Cahn}^h$ and $ C\BM^h$ are determined by the expressions:
 \be
  C_{Cahn}^h&=&\frac{2\int dz_h \,A^h_{Cahn}(z_h)}{\int dz_h \, [D_{u_V}^h(z_h]}
=-\avk\, \frac{\int dz_h\,z_h[D_{q_V}^{h}(z_h)]/\sqrt{\avPT}}{\int dz_h\, [D_{q_V}^{h}(z_h)]},\\
  C\BM^h&=&\frac{2\int dz_h\, A^h\BM (z_h)}{\int dz_h \,[D_{u_V}^h(z_h]}\\
  &=&\frac{e\,\avk\BM^2\,\avp\C^2}{2\,M\BM\,M\C\avk\avp}\;\frac{\int dz_h\,[z_h^2\avk\BM  +2\avp\C]\,
 [\Delta^N D_{{q_V}\!\uparrow}^{h}(z_h)]\,
 /\avPT\BM ^{3/2}}{\int dz_h\,[D_{u_V}^{h}(z_h)]}\label{BM22}\\\nn
 \avPT &=&\avp +z_h^2 \avk ,\quad \avPT\BM =\avp\C +z_h^2 \avk\BM .
 \ee

%%%%%%%%%%%%%%%%%%%%%%%%%%%%%%%%%%%%%%%%%%%%%%%%%%%%%%%%%%%%%%%%%%%%%%%%%%%%%%%%%%%%%%%%%%%%%%%%%%%%%%%%%%%%%%%%%%%%%%%%%%%%%

%%%%%%%%%%%%%%%%%%%%%%%%%%%%%%%%%%%%%%%%%%%%%%%%%%%%%%%%%%%%%%%%%%%%%%%%%%%%%%%%%%%%%%%%%%%%%%%%%%%%%%%%%%%%
\section*{Appendix B: The $A_{UU}^{\cos 2\phi_h}$ asymmetry}

 The structure function  $F_{UU}^{\cos 2\phi_h}$ that determines the azimuthal $A_{UU}^{\cos 2\phi_h}$ asymmetry, Eq. (\ref{Acos2}),
 has two  contributions - the leading twists-2 contribution from the BM functions and the twists-4 contribution of $1/Q^2$-order from Cahn
 effect:
\be F_{UU}^{\cos 2\phi_h} = F_{Cahn}^{\cos 2\phi_h}+ F_{BM}^{\cos 2\phi_h}
\ee
Again we shall consider only difference cross sections  $(h-\bar h)$ on deuteron
target. In this case it is
 only the sum of the valence-quark parton densities $Q_V$ that enter these functions.

 For  BM contribution on deuteron target for $h=\pi^+,K^+,h^+$ we have
~\cite{we0}:
\be
 F_{BM,d}^{\cos 2\phi_h,h-\bar h} (\xb ,z_h,Q^2,P_T^2)={\cal A}^{\cos 2\phi_h}\BM (z_h,P_T^2)\,\Delta f\BM^{Q_V} (\xb ,Q^2)
  [\Delta^N D^h_{q_V\uparrow } (z_h,Q^2)]
  \ee
  where the flavour and hadron independent function  ${\cal A}^{\cos 2\phi_h}\BM (z_h,P_T^2)$ reads:
\be
{\cal A}^{\cos 2\phi_h}\BM (z_h,P_T^2)=-e\,K\,\frac{P_T^2}{M\BM M\C}\,\frac{e^{-P_T^2/\avPT\BM }}{\pi \avPT^3\BM}\,z_h,
\ee
$K$ is determined in Eq. (\ref{K}).

Performing the integration over $P_T$ and using the standard parametrization Eq.
(\ref{BM-Siv_dist1}) for the BM function, we obtain:
\be
\int dP_T^2\,F_{UU,d}^{\cos 2\phi_h,h-\bar h}&=&2\,\hat A^{h}\BM (z_h,Q^2) {\cal N}^{Q_V}\BM (\xb ) Q_V (\xb ,Q^2)\label{BM2}
\ee
where
\be
\hat A^{h}\BM (z_h,Q^2) &=&[\Delta^N D_{u_V\uparrow}^h(z_h,Q^2)]\int dP_T^2\,{\cal A}\BM^{\cos 2\phi_h}(z_h, P_T^2)\nn
&=&\frac{-\,e\,K\,z_h}{\pi M\BM
M\C\avPT\BM}\left[\Delta
^ND_{u_V\uparrow}^h(z_h,Q^2)\right].
\ee
Eq. (\ref{BM2}) implies that if we can neglect $Q^2$-dependencies in $Q_V$ and in the FF the $\xb$- and $z_h$-dependencies  will factorize,
 $[\Delta^N D_{u_V\uparrow}^h]$ is given in Eq. (\ref{D2}).

The Cahn contribution to the asymmetry looks more complicated as the integration over $k_\perp$ that comes from the convolution of the TMD PDFs and
FFs cannot be fulfilled analytically. Nevertheless it has the same structure:
\be
 F_{Cahn,d}^{\cos 2\phi_h, h-\bar h} (\xb ,z_h,Q^2,{\bf P}_T)&=&\frac{2}{Q^2}\,Q_V(\xb ,Q^2)\left [D_{q_V}^{h}(z_h,Q^2)\right]\nn
 &&\times\int d^2{\bf k_\perp}\,d^2{\bf p_\perp}\,\left(2(\hat{\bf P_T}\cdot {\bf k_\perp} )^2-k_\perp^2\right)\; \frac{e^{-k_\perp^2/\avk}}{\pi \avk}\;\frac{e^{-k_\perp^2/\avk}}{\pi \avk}\,\delta^2({\bf P_T}-z_h{\bf k_\perp}-{\bf p_\perp})\nn
 &=&\frac{2}{Q^2}\,Q_V(\xb ,Q^2)\left [D_{q_V}^{h}(z_h,Q^2)\right]\,\frac{1}{2\pi^2\avk \avp}\; I(z_h,P_T^2)
 \ee
 where $[D_{q_V}^{h}]$ is given in Eq. (\ref{D1}),  and
 \be
 I=e^{-\frac{P_T^2}{\avp}}\int dk_\perp^2 k_\perp^2 e^{-k_\perp^2\frac{\langle P_T^2\rangle }{\avk\avp}}
 \int_0^{2\pi} d\phi\,\cos 2\phi \,e^{a\cos\phi},\quad a=\frac{2z_hk_\perp P_T}{\avp}.
 \ee
 For the integrated over ${\bf P}_T$  contribution of the Cahn effect we obtain:
 \be
 \int d^2{\bf P}_T F_{Cahn,d}^{\cos 2 \phi_h,h-\bar h} = \frac{2}{Q^2}\,\hat A_{Cahn}^{h}(z_h,Q^2) Q_V(\xb ,Q^2)\label{Cahn2}
 \ee
 where
\be
\hat A_{Cahn}^{h}(z_h,Q^2)\equiv \frac{1}{2\pi\avk \avp}\; J(z_h)[D_{q_V}^h(z_h,Q^2)],\qquad J(z_h)\equiv \int dP_T^2\,I(z_h,P_T^2)
\ee

>From Eqs. (\ref{BM2}) and (\ref{Cahn2}), and using Eq. (\ref{UU}),  we obtain the following
expression for the asymmetry $A_{UU,d}^{\cos 2\phi_h,h-\bar h}$:
\be
A_{UU,d}^{\cos 2\phi_h,h-\bar h }(\xb ,z_h,Q^2)=
\frac{2 [(1-y)/Q^4]\,Q_V(\xb ,Q^2)
\left\{\hat A\BM^h (z_h,Q^2){\cal N}\BM^{Q_V}(\xb )+[1/Q^2]\,\hat A_{Cahn}^h (z_h,Q^2)\right\}}{[[1+(1-y)^2]/Q^4]\,Q_V(\xb ,Q^2)[D_{u_V}^h(z_h,Q^2)]}
\ee

Neglecting $Q^2$-dependence in the $\xb$-bins in $Q_V$, the valence quark densities $Q_V$ in the nominator and in the denominator cancel out.
Neglecting further, the $Q^2$-dependence in the FFs and integrating over $z_h$
for the $\xb$-dependent $\cos 2\phi_h$-asymmetry we obtain:
\be
A_{UU,d}^{\cos 2\phi_h,h - \bar h}(\xb )=\hat \Phi (\xb)\,\left\{\hat C\BM ^h\,{\cal N}\BM^{Q_V}(\xb )+\frac{MM_d}{Q^2}\,\hat C_{Cahn}^h\right\}
\label{cos2_2}\ee
where
\be
\hat \Phi (\xb)&=&\frac{2\,(1-\bar y)}{[1+(1-\bar y)^2]}\nn
\hat C\BM ^h&=&\frac{-e\,K}{M\BM M\C}\,\frac{\int dz_h\,z_h\,[\Delta^N D_{u_V\uparrow}^h(z_h)]/
\avPT\BM}{\int dz_h\,[ D_{u_V}^h(z_h)]}\nn
\hat C_{Cahn}^h&=& \frac{1}{2\,MM_d\avk \avp}\;\frac{\int dz_h \,[ D_{u_V}^h(z_h)] \int dP_T^2\,I(z_h,P_T^2)}{\int dz_h\,[ D_{u_V}^h(z_h)]}
\ee
where $\bar y$ is given in Eq. (\ref{y}). Eq. (\ref{cos2_2}) is exactly our Eq. (\ref{Acos2phi}).

%%%%%%%%%%%%%%%%%%%%%%%%%%%%%%%%%%%%%%%%%%%%%%%%%%%%%%%%%%%%%%%%%%%%%%%%%%%%%%%%%%%%%%%%%%%%%%%%%%%%%%%%%%%%%%%%%%%%%%%%%

 \end{document}